\documentclass[fleqn,12pt]{article}
\usepackage{amsmath}
\usepackage{amsfonts,amssymb,latexsym}
\usepackage[vcentermath,enableskew]{youngtab}
\usepackage[nosort]{cite}
\usepackage[hyperref,bulletsep]{collect}
\def\baselinestretch{1.1}
\newcommand{\ft}[2]{{\textstyle\frac{#1}{#2}}}

\def\1bar{1\hskip -.275cm -}
\def\2bar{2\hskip -.275cm -}
\def\3bar{3\hskip -.275cm -}

\newsavebox{\uuunit}

\newcommand{\mis}{\!-\!}

\newcommand{\mathon}{\mathversion{bold}}
\newcommand{\mathoff}{\mathversion{normal}}
\newcommand{\la}{\label}

\newcommand{\irreps}{representations}
\newcommand{\be}{\begin{equation}} \newcommand{\ee}{\end{equation}}
\newcommand{\bea}{\begin{eqnarray}} \newcommand{\eea}{\end{eqnarray}}
\newcommand{\ben}{\begin{displaymath}}
\newcommand{\een}{\end{displaymath}}
\newcommand{\dalpha}{\dot{\alpha}} \newcommand{\dbeta}{\dot{\beta}}
\newcommand{\nn}{\nonumber} \newcommand{\non}{\nonumber\\}

\newcommand{\jb}{\bar{\jmath}}

\newcommand{\hs}{\mathfrak{hs}(2,2|4)}
\newcommand{\psu}{\mathfrak{psu}(2,2|4)}

\makeatletter
\let\old@makecaption=\@makecaption
\def\@makecaption{\small\old@makecaption}
\makeatother

\makeatletter \@addtoreset{equation}{section} \makeatother

\makeatletter
\let\old@startsection=\@startsection
\renewcommand{\@startsection}[6]{\old@startsection{#1}{#2}{#3}{#4}{#5}{#6\mathversion{bold}}}
\makeatother

\newcommand{\hypref}[2]{\ifx\href\asklfhas #2\else\href{#1}{#2}\fi}

\newcommand{\tabref}[1]{Tab.~\ref{#1}}

\newcommand{\sfrac}[2]{{\textstyle\frac{#1}{#2}}}
\newcommand{\half}{\sfrac{1}{2}}

\newcommand{\tr}{{\rm Tr}\,}
\newcommand{\alg}[1]{\mathfrak{#1}}

\newcommand{\alU}{\alg{u}}
\newcommand{\alSU}{\alg{su}}

\newcommand{\alSL}{\alg{sl}}
\newcommand{\alSO}{\alg{so}}

\newcommand{\alPSU}{\alg{psu}}
\newcommand{\alhs}{\alg{hs}}
\newcommand{\mult}{\mathcal{V}}

\newcommand{\order}[1]{\mathcal{O}(#1)}

\newcommand{\superN}{\mathcal{N}}
\newcommand{\gym}{g_{\scriptscriptstyle\mathrm{YM}}}

\newcommand{\cder}{\mathcal{D}}

\newcommand{\state}[1]{|#1\rangle}%
\newcommand{\bigbrk}[1]{\bigl(#1\bigr)}

\newcommand{\nln}{\nonumber\\}
\newcommand{\nl}{\nonumber\\&&\mathord{}}

\newcommand{\earel}[1]{\mathrel{}&#1&\mathrel{}}
\newcommand{\eq}{\earel{=}}
\newenvironment{myeqnarray}{\arraycolsep0pt\begin{eqnarray}}{\end{eqnarray}\ignorespacesafterend}
\newenvironment{myeqnarray*}{\arraycolsep0pt\begin{eqnarray*}}{\end{eqnarray*}\ignorespacesafterend}

\def\[{\begin{equation}}
\def\]{\end{equation}}
\def\<{\begin{myeqnarray}}
\def\>{\end{myeqnarray}}

\ifx\href\asklfhas\newcommand{\href}[2]{#2}\fi
\newcommand{\arxivno}[1]{\href{http://arxiv.org/abs/#1}{#1}}

\newdimen\squaresize \squaresize=12pt
\newdimen\thickness \thickness=0.7pt

\def\square#1{\hbox{\vrule width \thickness
   \vbox to \squaresize{\hrule height \thickness\vss
      \hbox to \squaresize{\hss#1\hss}
   \vss\hrule height\thickness}
\unskip\vrule width \thickness} \kern-\thickness}

\def\cut#1{\hbox{\vrule width-1 \thickness
   \vbox to \squaresize{\hrule height-1 \thickness\vss
      \hbox to \squaresize{\hss#1\hss}
   \vss\hrule height-1\thickness}
\unskip\vrule width +4 \thickness} \kern-\thickness}

\def\vsquare#1{\vbox{\square{$#1$}}\kern-\thickness}

\def\young#1{
\vbox{\smallskip\offinterlineskip \halign{&\vsquare{##}\cr #1}}}

\newcommand{\tinyyoung}[1]{
\squaresize=7pt \thickness=0.4pt \mbox{\tiny\young{#1}}
\squaresize=12pt \thickness=0.7pt}

\begin{document}

\thispagestyle{empty}

\begin{center}

{\small\ttfamily AEI~2004-025\hspace*{0.8cm}
ROM2F/04/08\hspace*{0.8cm} DESY 04-058\hspace*{0.8cm}
\arxivno{hep-th/0405057}%upload it as it stands, it's fine
}

\end{center}

%\bigskip

\begin{center}

\renewcommand{\thefootnote}{\fnsymbol{footnote}}

{\mathon\bf\Large  Higher Spin Symmetry
and $\mathcal{N}=4$ SYM\par \mathoff}%
\bigskip\bigskip

\addtocounter{footnote}{1} \textbf{
N.~Beisert\footnote{\texttt{nbeisert@aei.mpg.de}},
M.~Bianchi\footnote{\texttt{Massimo.Bianchi@roma2.infn.it}},
J.~F.~Morales\footnote{\texttt{Francisco.Morales@lnf.infn.it}},
and H.~Samtleben\footnote{\texttt{Henning.Samtleben@desy.de}}}

\addtocounter{footnote}{-3} \vspace{.3cm} $^\fnsymbol{footnote}$
\textit{ Max-Planck-Institut f\"ur Gravitationsphysik\\
Albert-Einstein-Institut\\
Am M\"uhlenberg 1, D-14476 Potsdam, Germany}\par

\addtocounter{footnote}{1} \vspace{.3cm} $^\fnsymbol{footnote}$
\textit{
Dipartimento di Fisica and INFN \\
Universit\`a di Roma ``Tor Vergata''\\
00133 Rome, Italy}\par

\addtocounter{footnote}{1} \vspace{.3cm} $^\fnsymbol{footnote}$
\textit{
Dipartimento di Fisica, Universit\`a di Torino,\\
Laboratori Nazionali di Frascati,\\
Via E. Fermi, 40, I-00044 Frascati (Rome), Italy}\par

\addtocounter{footnote}{1} \vspace{.3cm} $^\fnsymbol{footnote}$
\textit{
II. Institut f\"ur Theoretische Physik, Universit\"at Hamburg,\\
 Luruper Chausse 149, D-22761 Hamburg, Germany}\par

\setcounter{footnote}{0}

\end{center}

\begin{abstract}
We assemble the spectrum of single-trace operators in free
$\mathcal{N}= 4$ $SU(N)$ SYM theory into irreducible
representations of the Higher Spin symmetry algebra $\hs$. Higher
Spin representations or \emph{YT-pletons} are associated to Young
tableaux (YT) corresponding to representations of the symmetric
group compatible with the cyclicity of color traces.  After
turning on interactions $g_{\scriptscriptstyle\mathrm{YM}}\neq 0$,
YT-pletons decompose into infinite towers of representations of
the superconformal algebra $\psu$ and anomalous dimensions are
generated. We work out the decompositions of tripletons with
respect to the $\mathcal{N}=4$ superconformal algebra $\psu$ and
compute their anomalous dimensions to lowest non-trivial order in
$g_{\scriptscriptstyle\mathrm{YM}}^2 N$ at large $N$. We then
focus on operators/states sitting in semishort multiplets of
$\psu$. By passing them through a semishort-sieve that removes
superdescendants, we derive compact expressions for the partition
function of semishort primaries.

\end{abstract}

\def\baselinestretch{1.1}

\newpage

\section{Introduction}

Pushing Maldacena's conjecture \cite{Maldacena:1998re} to its
extreme consequences, one is led to conclude that free ${\cal
N}=4$ super Yang-Mills theory (SYM) with $SU(N)$ gauge group
should be holographically dual to type IIB superstring on an
extremely curved $AdS_5 \times S^5$
\cite{Sundborg:1999ue,Sundborg:2000wp,Haggi-Mani:2000ru,WittenJH60}.
The Hagedorn growth of single-trace gauge-invariant SYM operators
at large $N$ precisely reproduces (tree level) string expectations
and one is led to take the limit seriously
\cite{Sundborg:1999ue,Sundborg:2000wp,Haggi-Mani:2000ru,WittenJH60,Mikhailov:2002bp,Polyakov:2001af,Aharony:2003sx}
and try to match the spectra on the two sides of the
correspondence.
 At vanishing gauge coupling constant ($\gym=0$) ${\cal N}=4$
 SYM theory develops a higher
spin (HS) symmetry $\hs$. One thus expects the same should happen
in the tensionless limit
\cite{Isberg:1992ia,Isberg:1994av,Bredthauer:2004kv,Lindstrom:2003mg,Bonelli:2003zu,Bakas:2004jq}
or rather at some very small radius of order $R\approx
\sqrt{\alpha^\prime}$
\cite{Sezgin:2002rt,Dhar:2003fi,Gopakumar:2003ns,Gopakumar:2004qb,deMedeiros:2003hr,Son:2003zv}
for the type IIB superstring on $AdS_5 \times S^5$ (see
\cite{Konstein:2000bi,Francia:2002aa,Vasiliev:2003ev,Sagnotti:2003qa,Vasiliev:2004qz}
for studies of higher spin gauge theories in various dimensions).
Despite some progress
\cite{Berkovits:1999xv,Berkovits:1999im,Berkovits:2002zv}, string
quantization in the presence of RR backgrounds is poorly
understood in general, let alone at large curvatures, and one has
to devise some alternative strategy for the time being.

In \cite{Bianchi:2003wx} a precision test of the correspondence
was carried out by first extrapolating the naive Kaluza-Klein (KK)
reduction on $S^5$ of the type IIB superstring spectrum from ten
dimensions to the point of enhanced HS symmetry and then
postulating a mass formula for the resulting string excitations
that could account for the appearance of the expected massless HS
gauge fields. The impressive agreement with single-trace SYM
operators at large $N$ up to dimension 4, including those
belonging to genuinely long supermultiplets, led us to suspect
that one could do better and find a more accurate energy formula
valid for all superstring states at the point of HS symmetry
enhancement. Indeed by relying on the BMN limit
\cite{Berenstein:2002jq} and extrapolating the plane-wave
frequencies down to finite $J$ (at $\gym = 0$!) such a formula was
found \cite{Beisert:2003te}
\[
\Delta = J + \nu \;, \label{magic}
\]
where $\nu = \sum_n N_n$ is the string occupation number and $J$
the charge under an $\alSO(2)$ subgroup of a hidden $\alSO(10)$
symmetry \cite{Beisert:2003te,Bars:1995uh} that organizes the KK
string spectrum \footnote{While this paper was being published an
interesting paper by Itzhak Bars appeared on the archive
\cite{Bars:2004dg} that discussed how the spectrum of higher spin
currents in N=4 SYM could be related to a particular gauge fixing
of the two-time superstring and reviewed previous work
\cite{Bars:1995uh} where the higher dimensional origin of the
SO(10) symmetry had been advocated.} . Despite its simplicity,
(\ref{magic}) encompasses the correct `energies' for all string
states to match those of SYM operators up to dimension $\Delta=10$
together with their superdescendants that neatly assemble into
(long) supermultiplets of $\psu$.\footnote{The upper bound of
$\Delta=10$ is imposed on us by computer capabilities.}

The aim of this paper is to rearrange the SYM / string spectrum
into multiplets of the higher spin (HS) extension $\hs$ of the
superconformal group. Representations of $\hs$ can be built out of
multiple tensor products of singleton multiplets. The singleton of
$\hs$ turns out to coincide with the singleton of $\psu$ with
vanishing central charge, that consists of the fundamental SYM
fields together with their derivatives
\cite{Vasiliev:2004qz,Sezgin:2002rt,Sezgin:2001zs,Sezgin:2001yf}.
In the absence of abelian factors in the gauge group, the
singleton does not give rise to well defined scaling operators. In
the holographic description it corresponds to the low-lying
open-string excitations that cannot propagate in the bulk of
$AdS$. Its (gauge invariant) composites which correspond to
closed-string excitations
can~\cite{Gunaydin:1985fk,Ferrara:1998ej,Ferrara:1999ed}. The
symmetric product of two singletons gives rise to the HS
`massless' doubleton containing all twist 2 gauge-invariant
operators and their superpartners. They are dual to the HS gauge
fields and their superpartners in the bulk. More precisely, the
symmetric doubleton product decomposes into an infinite number of
$\alSU(2,2|4)$ multiplets
\cite{Ferrara:1998pr,Sezgin:2001zs,Sezgin:2001yf,Sezgin:2002rt,Dolan:2001tt,Dolan:2002zh,Dobrev:2004ha}
\[
({\cal V}_F \times {\cal V}_F)_S ~=~ \sum_{n=0}^\infty {\cal
V}_{2n} \;. \label{dupro}
\]
Here ${\cal V}_F$ denotes the singleton and ${\cal V}_{2n\geq 2}$
are semishort current multiplets with primaries transforming as
singlets of $SU(4)$ and carrying spin $2n-2$. For instance ${\cal
V}_{2}$ denotes the (semishort) ${\cal N} =4$ Konishi multiplet
\cite{Andrianopoli:1998ut,Ferrara:1999zg}. Finally, ${\cal V}_0$
is the $\ft12$-BPS supercurrent multiplet. The anti-symmetric
doubletons with odd spin ${\cal V}_{2n+1}$ do not appear in the
free SYM spectrum due to the cyclicity of the trace but play a
role in the interactions.

As we will see, the pattern persists for higher tensor multiplets.
For gauge group $SU(N)$, the tensor product of $L$ singletons
decomposes into \irreps of $\hs$, that may be termed {\em
YT-pletons} since they are completely classified by those Young
tableaux (YT) with $L$ boxes that are compatible with the
cyclicity of the trace.  At large $N$, mixing among single and
multi trace operators is suppressed
\cite{Bianchi:2002rw,Beisert:2002bb,Constable:2002vq,Arutyunov:2002rs,Bianchi:2003eg}.
It is impossible anyway at $L=3$, where we will find only two
`massive' \irreps: the totally symmetric one including the first
KK recurrence of the HS `massless' doubleton, and the totally
antisymmetric one, always present, that includes part of the lower
spin Goldstone / St\"uckelberg fields needed for the Higgsing of
the HS `massless' gauge fields when departing from the HS symmetry
enhancement point
\cite{Sezgin:2001zs,Sezgin:2001yf,Sezgin:2002rt,Bianchi:2003wx}.
For other gauge groups, such as $SO(N)$ or $Sp(2N)$, one has to
also take into account the symmetry under transposition that is
holographically dual to a combination of worldsheet parity,
spacetime inversion and fermion parity
\cite{Bianchi:1992eu,Bianchi:1998rf,Witten:1998bs,Witten:1998xy}.
At $L=3$ this projects out the completely symmetric YT tableau
leaving only the completely antisymmetric ``Goldstone'' multiplet.
Related aspects of the SYM spectrum have been studied in
\cite{Heslop:2001dr,DHoker:2003vf,Heslop:2003xu,Aharony:2003sx}.

In the boundary theory, turning on interactions ($\gym\neq 0$)
breaks the HS symmetry down to the superconformal supergroup
$\alPSU(2,2|4)$. As a result, both massless and massive
representations of $\hs$ typically decompose into infinite series
of $\psu$ supermultiplets. Massive representations of HS symmetry
algebras have not been much studied in the past
\cite{Metsaev:2003cu,Vasiliev:2004cm}. Here we present the
simplest occurrences of massive representations of $\hs$. They
will play a crucial role in the Pantagruelic Higgs mechanism
(`Grande Bouffe') in the AdS bulk that gives masses to all HS
gauge fields except the graviton and its superpartners.

In a (superconformal) quantum field theory the violation of a
symmetry due to quantum effects reflects into the corresponding
current acquiring an anomalous dimension
\cite{Anselmi:1997dd,Anselmi:1998ys,
Anselmi:1998am,Anselmi:1998ms,Anselmi:1998bh,Anselmi:1999bb}.
Anomalous dimensions can be computed either by {\it old-fashion}
QFT methods
\cite{Bianchi:1999ge,Bianchi:2000hn,Bianchi:2001cm,Bianchi:2002rw,
Arutyunov:2002rs,Gross:2002su,Santambrogio:2002sb,Eden:2003sj,Kovacs:2003rt,Dolan:2001tt,Dolan:2002zh}
or by {\it brand-new} techniques based on the identification of
the planar dilatation operator \cite{Beisert:2003tq} with the
Hamiltonian of an integrable super-spin chain
\cite{Minahan:2002ve,Beisert:2003jj,Beisert:2003yb,Beisert:2004hm}.
Non planar dynamics is described by a spin chain with non-local
interactions accounting for joining and splitting of SYM traces
\cite{Beisert:2003tq,Bellucci:2004ru}. Here we apply the spin
chain techniques to determine the one-loop anomalous dimensions
for some operators consisting of three constituent fields in
${\cal N}=4$ SYM.

The plan of the paper is as follows.  In section~\ref{shsa} we
discuss the algebra $\hs$, the HS extension of the superconformal
algebra $\psu$, and its representations. We consider first the
case of $\alhs(1,1)$, the HS extension of $\alSU(1,1)\sim
\alSL(2)$, spanned in ${\cal N}=4$ SYM by a single (complex)
scalar field and its derivatives in a given (complex) direction.
This truncation illustrates already the main features of the HS
representation theory which apply to HS extensions of Poincar{\'e}
and superconformal algebras in any dimension.\footnote{HS algebras
in other dimensions are supported by non-conformal free SYM
theories living on D$p$-branes are their gravity duals on warped
AdS
geometries~\cite{Itzhaki:1998dd,Gherghetta:2001iv,Morales:2002ys,Morales:2002uh}.}
In particular we argue that irreducible representations of
 $\hs$ are in one-to-one correspondence with Young tableaux built out of
 singletons. Only a subset of these representations survives tracing over
 color indices.

In section~\ref{smult} we describe how the spectrum of operators
in free ${\cal N}=4$ SYM can be assembled into irreducible
representation of $\hs$. We determine via Polya theory the set of
Young tableaux surviving the tracing over $SU(N)$ gauge indices
 and display the
$\psu$ content of the first occurrences of massive HS
representations at `twist' 3. Higher $L$-pletons involve more
complicated decompositions spanning several infinite towers of
${\cal N}=4$ multiplets.

In section~\ref{ssemishort} we restrict to states sitting in
semishort multiplets of the ${\cal N}=4$ SCA. Disposing of
superdescendants by means of a semishort-sieve, we derive compact
expressions for ${\cal Z}^{\rm short}_{\rm suprim}$, the partition
function of BPS and semishort primaries. In section~\ref{sanodim}
we turn on interactions, i.e.~a small non-vanishing SYM coupling
$\gym\neq 0$, that break $\hs$ down to $\psu$ and compute the
anomalous dimensions of tripletons to lowest non-trivial order in
$\gym^2 N$ at large $N$. Finally, in section \ref{sconclusions},
we conclude with some comments on $L$-pletons and integrability.
Appendix~\ref{sa1} introduces a unifying notation for ${\cal N}=4$
UIR's and shortenings originally discussed in \cite{Dobrev}.
Appendices~\ref{sao} and~\ref{a123} collect other useful formulae.

\section{The higher spin algebra and its representations}
\label{shsa}

At vanishing gauge coupling constant ($\gym=0$) the SCA
$\alPSU(2,2|4)$ of ${\cal N}=4$ SYM theory gets enhanced to the HS
symmetry  algebra
$\alhs(2,2|4)$~\cite{Sezgin:2001zs,Sezgin:2001yf,Sezgin:2002rt,
Vasiliev:2003ev,Alkalaev:2002rq,Vasiliev:2001wa,Anselmi:1998ms,Anselmi:1998bh,Anselmi:1999bb}.
The HS symmetry algebra is generated by an infinite set of
conserved currents of arbitrarily high (even) spin $s=2n$
associated to totally symmetric and traceless tensors
 \[
 {\cal J}_{\mu_1\ldots \mu_{2n}}=\tr \varphi^i \partial_{(\mu_1}\ldots
 \partial_{\mu_{2n})} \varphi_i+\ldots\;,
\]
and their superpartners. Together with the lowest ultra-short
$\ft12$-BPS multiplet that contains the unbroken currents of the
superconformal algebra $\alPSU(2,2|4)$, the infinite tower of HS
multiplets builds a single {\it massless} multiplet of the HS
algebra $\alhs(2,2|4)$, the {\it doubleton}~(\ref{dupro}). The
doubleton collects all gauge-invariant operators built from two
SYM elementary fields $\{A_\mu, \lambda_A^\alpha,
\bar\lambda^A_{\dot\alpha}, \varphi^i\}$ and derivatives thereof
modulo field equations. In general, all states belonging to a HS
multiplet have a common length $L$, i.e.~number of constituents or
`partons' ($L=2$ for the doubleton), since any linearly realized
symmetry at $\gym=0$ preserves the number of letters.\footnote{A
closely related notion is the `twist' $\tau=\Delta - s$, with
$\Delta$ the scaling dimension and $s$ the spin. $L=\tau$ for
semishort primaries.} However, for a given length $L>2$, several
HS multiplets appear. In \cite{Sezgin:2001zs}, the $\alPSU(2,2|4)$
content of the HS massless multiplet was determined, and the HS
gauge theory realizing the algebra $\alhs(2,2|4)$ on $AdS_5$ was
formulated at the linearized level.

Here we consider {\it massive} representations\footnote{At
$\gym=0$, the lowest cases ($L=3,4$) still contain some marginal
scalar operators dual to some massless scalars in bulk.} of
$\alhs(2,2|4)$. As we will see they play a crucial role in the
decomposition of the free ${\cal N}=4$ SYM spectrum on the
boundary and in the Pantagruelic Higgs mechanism in the AdS bulk.
Although our discussion will focus on $\alhs(2,2|4)$ for its
relevance to ${\cal N}=4$, the analysis can be adapted to HS
extensions of superconformal or Poincar{\'e} groups in other
dimensions. To this purpose, it is instructive to start describing
representations of the simplest HS algebra $\alhs(1,1)$, the HS
extension of $\alSU(1,1)\sim \alSL(2)$ \cite{Vasiliev:2004qz}.
This subalgebra, generated by a single derivative, is part of any
HS algebra independently of the dimension. As we will see later
on, the discussion of $\alhs(2,2|4)$, like any other HS extension,
is an almost straightforward generalization of this simple case.

The algebra $\alhs(2,2|4)$ and its irreducible representations, or
{\it YT-pletons}, are subsequently discussed in subsection
\ref{afirst}. The decomposition of YT-pletons into an infinite sum
of irreducible representations of the ${\cal N}=4$ superconformal
subalgebra $\psu$ will be further discussed in
section~\ref{smult}.

\subsection{Representations of $\alhs(1,1)$}

Here we describe the representations of the HS algebra
$\alhs(1,1)$. This algebra is realized on a single complex scalar
(in the adjoint of the gauge group) and its derivatives along a
chosen (complex) direction. The aim of this section is to
establish a one-to-one correspondence between irreducible
representations of $\alhs(1,1)$ and Young tableaux made out of
singletons. This section can be read independently of the rest of
the paper and applies to HS algebras in any dimension containing
$\alhs(1,1)$ as a subgroup.

\subsection*{$\alSL(2)$}

We start by considering the $\alSL(2)$ subalgebra:
\[
 [ J_-,J_+ ]=2\,J_3\;, \qquad
 [J_3,J_{\pm} ]=\pm J_{\pm} \;.
 \]
 This algebra may be represented in terms of oscillators
\[
 J_+=a^\dagger + a^\dagger a^\dagger a\;, \qquad
 J_3= \ft12+a^\dagger a\;, \qquad
 J_-=a\;,
\label{sl2g} \] where, as usual, $[a,a^\dagger]=1$. For our
purpose it is convenient to work in the space of functions
$f(a^\dagger)$, wherein $a={\partial/\partial a^\dagger}$.
$\alSL(2)$ highest weight states (HWS's) are defined by
 \[
 J_- f(a^\dagger)|0\rangle =0  \quad \Rightarrow  \quad f(a^\dagger)=1
 \;.
 \]
 Any state $(a^\dagger)^n|0\rangle$ in this defining representation
 can be generated from its HWS $|0\rangle$ by acting with $J_+^n$. Therefore
$f(a^\dagger)$ defines a single irreducible representation of
$\alSL(2)$. For later purposes, we call this representation {\it
singleton} and denote it by~$V_F$. The $\alSL(2)$ spin of this
representation is $-J_3 \,|0\rangle =-\ft12 |0\rangle$.   In
${\cal N}=4$ SYM, the components of the $\alSL(2)$ singleton may
be chosen in $\alPSU(2,2|4)/\alSL(2)$ different ways. In
particular, the HWS can be identified with the (complex) scalar
$Z=\varphi^5 + i \varphi^6$ and its $\alSL(2)$ descendants can be
generated by the action of the derivative along a chosen complex
direction ${\cal D}=D_1 + i D_2$,
\[
(a^\dagger)^n |0\rangle \quad \leftrightarrow \quad {\cal D}^n Z
\;.
\]
In a similar way, the tensor product of $L$ singletons may be
represented in the space of functions
$f(a^\dagger_{(1)},a^\dagger_{(2)},\ldots,a^\dagger_{(L)})$ with
$a^\dagger_{(s)}$ acting on the $s^{\rm th}$ site. The resulting
representation is no longer irreducible. This can be seen by
looking for $\alSL(2)$ HWS's
\[
J_-\, f(a^\dagger_{(1)},\ldots,a^\dagger_{(L)})=\sum_{s=1}^L
\,\partial_s \,f(a^\dagger_{(1)},\ldots,a^\dagger_{(L)})=0 \;.
\]
with $\partial_s={\frac{\partial}{\partial a^\dagger_{(s)}}}$.
There are indeed several solutions to these equations given by all
possible functions of the differences
$f_L(a^\dagger_{(s)}-a^\dagger_{(s')})$. A basis for $\alSL(2)$
HWS's can be chosen as
\[
|j_{1},\ldots,j_{L-1}\rangle =
(a^\dagger_{(L)}-a^\dagger_{(1)})^{j_{1}}
(a^\dagger_{(L)}-a^\dagger_{(2)})^{j_{2}}\ldots
(a^\dagger_{(L)}-a^\dagger_{(L-1)})^{j_{L-1}}\,|0\rangle \;,
\label{hws}
\]
with spin $J_3=\ft12+\sum_{s} j_{s}$. In particular for $L=2$ one
finds the known result
\[
V_F\times V_F=\sum_{j=0}^\infty V_j \;,
\]
with $V_j$ generated by acting with $J_+$ on the HWS
$|j\rangle=(a^\dagger_{(2)}-a^\dagger_{(1)})^j\,|0\rangle$. The
corresponding states in free ${\cal N}=4$ SYM follow from the
dictionary \[ (a^\dagger_{(1)})^{n_{1}} \,(a^\dagger_{(2)})^{n_{2}
} \ldots (a^\dagger_{(L)})^{n_{L}}\,|0\rangle  \quad
\leftrightarrow \quad {\cal D}^{n_1} Z \,{\cal D}^{n_2}Z \ldots
{\cal D}^{n_L} Z\;,\] with $n_i\geq 0$.

\subsection*{$\alhs(1,1)$}

The HS extension $\alhs(1,1)$ of $\alSL(2)$ is defined by
introducing the HS generators \cite{Vasiliev:2004qz}
\[
 J_{p,q}= (a^\dagger)^{p} a^q\;.
\]
The $J_{p,q}$ clearly close under the commutator / product into an
HS algebra that contains the $\alSL(2)$ subalgebra~(\ref{sl2g}).
We call it $\alhs(1,1)$. The generators $J_{p,q}$ with $p<q$ are
raising operators. The singleton is again a representation of this
algebra since $|0\rangle$ is annihilated by all raising operators.
In the tensor product of $L$ singletons, HWS's of $\alhs(1,1)$
are solutions of
\begin{eqnarray}
 \sum_{i=1}^L \, (a^\dagger_{(i)})^{p}\,
 \partial_i^{q}\,f(a^\dagger_{(1)},..,a^\dagger_{(L)})=0\quad {\rm with}~p<q
 \;.
 \label{hseq}
\end{eqnarray}
Equations (\ref{hseq}) are highly restrictive and solutions are
rare. Our claim is that solutions to these equations,
i.e.~irreducible representations of $\alhs(1,1)$, are in
one-to-one correspondence with Young tableaux (YT) made out of $L$
boxes, i.e.~row increasing diagrams with boxes numbered always
increasingly along rows and columns (for a quick review on YT
decompositions see the Appendix of \cite{Bellucci:2004ru}).

Since $\alhs(1,1)$ HWS's are also HWS's of its $\alSL(2)$
subalgebra we can restrict our attention to states of the type
(\ref{hws}). For instance, for $L=2$, the condition $J_{0,1}
f(a^\dagger_{(1)},a^\dagger_{(2)})|0\rangle=0$ is solved by
\[
|j\rangle =(a^\dagger_{(2)}-a^\dagger_{(1)})^j|0\rangle \;.
\]
The conditions $J_{0,n}| j \rangle=0$ leave only $| 0 \rangle$ and
$| 1\rangle$ as solutions. Indeed the two states automatically
satisfy $J_{m\geq 1,p\geq 2}|j\rangle=0$ and therefore are HWS's.
They correspond to the two HWS's in the symmetric and
antisymmetric tensor product of two singletons respectively
\begin{eqnarray} &&\tinyyoung{1&2\cr}=|0\rangle
\quad\leftrightarrow\quad Z^2 \;,\nn\\
&&\tinyyoung{1\cr 2\cr}=(a^\dagger_{(2)}-a^\dagger_{(1)})|0\rangle
\quad\leftrightarrow\quad Z({\cal D}Z)- ({\cal D}Z)Z \;.
\end{eqnarray}
In ${\cal N}=4$ SYM, the antisymmetric doubleton is projected out
after tracing over gauge indices. The generalization to $L>2$
states is straightforward. The HWS for the
 completely symmetric representation
 is again given by $f(a^\dagger_{(1)},\ldots , a^\dagger_{(L)})=1$ while the HWS for
 the completely antisymmetric tableau can be
 written as a product of all the differences
\[\tinyyoung{\cr\cr \cr\cr}~=~
\prod^{L}_{i>j} (a^\dagger_{(i)}-a^\dagger_{(j)}) |0\rangle \quad
\leftrightarrow \quad Z \,({\cal D} Z) \,({\cal D}^2 Z) \ldots
({\cal D}^{L-1} Z)+{\rm antisymm.} \label{anti} \]
 That this state satisfies~(\ref{hseq}) can be seen by noticing that being completely antisymmetric,
 derivatives $\sum_i (a^\dagger_{(i)})^p\, \partial^q_i$ in the
 completely symmetric operator $J_{p,q}$
 cancel against each other.
  Similarly one can build more general solutions from tensoring $k$ columns of type (\ref{anti})
  leading to
 \[
\tinyyoung{& & &\cr &\cr &
\cr\cr}~=~\prod_{p=1}^k\,\prod^{L_p}_{i_p>j_p}
(a^\dagger_{(i_p)}-a^\dagger_{(j_p)}) |0\rangle \quad
\leftrightarrow \quad Z^{n_1} \,({\cal D} Z)^{n_2}  \ldots ({\cal
D}^{n_s} Z)+{\rm perms.} \] with $L=\sum_p\, L_p$, $n_i$ the
number of boxes in the $i^{\rm th}$ row and ``perms{}'' denoting
all permutations specified by the tableau. To each of these
solutions we associate a Young Tableaux with $k$ columns of length
$L_p$ and boxes labelled by $i_p\in \{ 1,2,\ldots , L_p\}$. We
believe that these are the only solutions to (\ref{hseq}) but we
have no rigorous proof of this uniqueness.

For example, HWS's for $L=3$ are given by
\begin{eqnarray}
&&\tinyyoung{1&2&3\cr}~=~|0\rangle   \quad \leftrightarrow \quad  Z^3 \;,\nn\\
&&\tinyyoung{1&2\cr
3\cr}~=~(a^\dagger_{(3)}-a^\dagger_{(1)})|0\rangle \quad
\leftrightarrow \quad   Z^2 ({\cal D}Z)-({\cal D}Z)Z^2 \;,\\
&&\tinyyoung{1&3\cr
2\cr}~=~(a^\dagger_{(2)}-a^\dagger_{(1)})|0\rangle\quad
\leftrightarrow \quad  Z ({\cal D}Z)Z-({\cal
D}Z)Z^2 \;,\nn\\
&&\tinyyoung{1\cr 2\cr
3\cr}~=~(a^\dagger_{(2)}-a^\dagger_{(1)})(a^\dagger_{(3)}-a^\dagger_{(1)})
(a^\dagger_{(3)}-a^\dagger_{(2)})|0\rangle \quad \leftrightarrow
\quad
 Z ({\cal D}Z) ({\cal D}^2 Z)  +{\rm antisymm.}\nn
\end{eqnarray}
In ${\cal N}=4$ SYM, the two HS 3-pleton multiplets associated to
the hooked tableaux $\tinyyoung{&\cr \cr}$ are projected out after
tracing over the gauge indices.

\subsection{A first look at $\alhs(2,2|4)$}
\label{afirst} In order to extend the previous analysis to the
case of our main interest, the higher spin algebra~$\alhs(2,2|4)$,
we need to recall some basic properties of this infinite
dimensional HS (super)algebra. To this end we closely follow
\cite{Sezgin:2001zs} and adopt their notations with minor changes.
The ${\cal N}=4$ superconformal algebra $\alPSU(2,2|4)$ can be
realized in terms of (super-)oscillators $\zeta_\Lambda = (y_a,
\theta_A)$ with:
\[
 {}[y_a, \bar{y}^b] = \delta_a{}^b{} \;,\qquad\qquad
\{\theta_A, \bar{\theta}^B\} = \delta^B{}_A \;,
\]
where $y_a, \bar{y}^b$ are bosonic oscillators with $a,b=1,...4$ a
Weyl spinor index of $\alSO(4,2)\sim \alSU(2,2)$ or, equivalently,
a Dirac spinor index of $\alSO(4,1)$, while
$\theta_A,\bar{\theta}^B$ are fermionic oscillators with
$A,B=1,...4$ a Weyl spinor index of $\alSO(6)\sim \alSU(4)$.

Generators of $\alPSU(2,2|4)$ are written as `traceless' bilinears
$\bar{\zeta}^\Sigma\zeta_\Lambda$ of superoscillators. In
particular, the `diagonal' combinations realize the compact
$\mathfrak{so}(6)$ and noncompact $\mathfrak{so}(4,2)$ bosonic
subalgebras respectively, while the mixed combinations generate
supersymmetries:
\begin{eqnarray} J^a{}_b &=& \bar{y}^a y_b - \ft12 K \delta^a{}_b
\;, \quad K = \ft12 \bar{y}^a y_a\;,
 \nn\\[1ex]
T^A{}_B &=& \bar{\theta}^A \theta_B - \ft12 B \delta^A_B \;, \quad
B = \ft12 \bar{\theta}^A \theta_A \;, \nn\\[1ex]
 {\cal Q}^A_a &=& \bar{\theta}^A
y_a\;,\qquad \qquad \quad   \bar{\cal Q}^a_A = \bar{y}^a \theta_A
\;. \label{supconfgen}
\end{eqnarray}
The combination
\[ C \equiv  K+B = \ft12 \bar{\zeta}^\Lambda \zeta_\Lambda\;,
\] commutes with all the remaining generators and is
thus a central element. The abelian ideal generated by $C$ can be
modded out e.g.~by setting  $C$ to zero. At least in perturbation
theory, this should make physical sense,  since the elementary SYM
fields $\{A_\mu, \lambda_A^\alpha, \bar\lambda^A_{\dot\alpha},
\varphi^i\}$ and their composites all have $C=0$.\footnote{In
principle, one can consider quotienting by $C - C_0$, where $C_0$
is any (half) integer. This would correspond to choosing as the
basic building block some singleton of $SU(2,2|4)$ with non
vanishing central charge $C=C_0$. These non self-conjugate
singletons play only a marginal accessory role in (perturbative)
${\cal N}=4$ SYM
theory\cite{Gunaydin:1998sw,Gunaydin:1998jc,Claus:1999xr,Fernando:2002wv,
Sezgin:2001zs,Sezgin:2001yf,Sezgin:2002rt,Andrianopoli:1999vr}. }
Finally, the combination $B$ is to be identified as the generator
of Intriligator's ``bonus
symmetry{}''~\cite{Intriligator:1998ig,Intriligator:1999ff} dual
to the `anomalous' $U(1)_B$ chiral symmetry of type IIB in the AdS
bulk. It acts as an external automorphism~\cite{Sezgin:2001zs}
that rotates the supercharges of the SCA. The  $\alPSU(2,2|4)$
invariant vacuum $|0\rangle$, annihilated by $\zeta_\Lambda$,
corresponds to the identity operator which can be viewed as the
trivial singlet representation.

The HS extension $\alhs(2,2|4)$ is roughly speaking generated by
odd powers of the above generators i.e.~combinations with equal
odd numbers of $\zeta_\Lambda$ and $\bar{\zeta}^\Lambda$. More
precisely, one first considers the enveloping algebra of
$\alPSU(2,2|4)$, which is an associative algebra and consists of
all powers of the generators, then restricts it to the odd part
which closes as a Lie algebra modulo the central charge $C$, and
finally quotients the ideal generated by $C$. It is easy to show
that $B$ is never generated in commutators (but $C$ is!) and thus
remains an external automorphism of $\alhs(2,2|4)$. Generators of
$\hs$ can be represented by `traceless' polynomials in the
superoscillators:
\begin{eqnarray}
\hs &=& \oplus_\ell\, {\cal A}_{2\ell+1} = \sum_{\ell=0}^\infty
\Big\{ {\cal J}_{2\ell+1}= P^{\Lambda_1\ldots
\Lambda_{2\ell+1}}_{\Sigma_1\ldots \Sigma_{2\ell+1}}\,
 \bar{\zeta}^{\Sigma_1}\!\dots\bar{\zeta}^{\Sigma_{2\ell+1}}\,
 \zeta_{\Lambda_1}\!\dots \zeta_{\Lambda_{2\ell+1}}
\Big\}\;,\label{phs}
\end{eqnarray}
with elements ${\cal J}_{2\ell+1}$ in ${\cal A}_{2\ell+1}$, where
$\ell$ is called the level, parametrized by traceless rank
$(2\ell\!+\!1)$ (graded) symmetric tensors $P^{\Lambda_1\ldots
\Lambda_{2\ell+1}}_{\Sigma_1\ldots \Sigma_{2\ell+1}}$. The
commutators of two elements however close only up to the ideal
generated by $C$. In
 particular they close on the subspace of physical states defined by the
 condition $C\equiv0$. The restriction to this subspace will be always understood.
Alternatively, the HS algebra can be more generally defined by
identifying generators differing by terms that involve $C$,
i.e.~${\cal J}\approx {\cal K}$ iff ${\cal J} - {\cal K}=\sum_
{k\ge 1} C^k {\cal H}_k$ \cite{Sezgin:2001zs}.

To each element in ${\cal A}_{2\ell+1}$ with $\alSU(2)_L\times
\alSU(2)_R$ spins $[j,\jb]$ is associated an $\alhs(2,2|4)$ HS
gauge field in the AdS bulk with labels $[j+\ft12,\jb+\ft12]$. The
$\alSU(4)\times \alSU(2)^2$ content of the HS currents can be
easily read off from (\ref{phs}) by expanding the polynomials in
powers of $\theta$'s up to 4, since $\theta^5=0$. There is a
single superconformal multiplet $\mult_{2\ell}$ at each level
$\ell\geq 2$. The lowest spin cases $\ell=0,1$, i.e.
$\hat{\mult}_{0,2}$, are special. They differ from the content of
doubleton multiplets $\mult_{0,2}$ by spin $s<1$ states
\cite{Sezgin:2001zs}. The content of (\ref{phs}) can then be
written as (tables 4,5 of \cite{Sezgin:2001zs})
\begin{eqnarray} \hat{\mult}_{0}&=& \big|\bar{{\bf
4}}_{[{\frac12},0]}+{\bf 1}_{[1,0]} \big|^2-
{\bf 1}_{[{\frac12},{\frac12}]}\nn\\
\hat{\mult}_{2}&=&  \big|{\bf 4}_{[{\frac12},0]}+{\bf 6}_{[1,0]}
+\bar{{\bf 4}}_{[{\frac32},0]}+{\bf 1}_{[2,0]} \big|^2\nn\\
 \mult_{2\ell}&=&
 \big| {\bf 1}_{[\ell-1,0]}+ {\bf 4}_{[\ell-{\frac12},0]}+
   {\bf 6}_{[\ell,0]}+ \bar{{\bf 4}}_{[\ell+{\frac12},0]}+
   {\bf 1}_{[\ell+1,0]} \big|^2 \;,   \qquad \ell\geq2\;,
\end{eqnarray} with ${\bf r}_{[j+{\frac12},\jb+{\frac12}]}$ denoting the
$\alSU(4)$ representation~${\bf r}$ and the labels of the
$\alU(1)^2\in \alSU(2)^2$ HWS's. Complex conjugates are given by
conjugating $\alSU(4)$ representations and exchanging the spins
$j\leftrightarrow \jb$. The product is understood in $\alSU(4)$
while $\alU(1)^2$ labels simply add.
 The highest spin state ${\bf 1}_{[\ell+1,\ell+1]}$ corresponds to the state
 $y^{2\ell+1}\bar{y}^{2\ell+1}$ with no $\theta$'s, ${\bf 4}_{[\ell+{\frac12},\ell+1]}$,
 $\bar{{\bf 4}}_{[\ell+1,\ell+{\frac12}]}$ to $y^{2\ell}\bar{y}^{2\ell+1} \theta^A$,
$y^{2\ell+1}\bar{y}^{2\ell} \bar{\theta}_A$, and so on. For
$\ell=0,1$, states with negative $j,\jb$ should be deleted. In
addition we subtract the current ${\bf 1}_{[{\frac12},{\frac12}]}$
at $\ell=0$ associated to $C$. In the ${\cal N}=4$ notation
introduced in Appendix~A,  ${\mult}_{2\ell}$ corresponds to the
semishort multiplet ${\mult}_{[000][\ell-1^*,\ell-1^*]}^{2\ell,0}$
(see also table \ref{hscont} in Appendix \ref{a123}).

\subsection*{Representations of $\alhs(2,2|4)$}

The basic representation of both $\alPSU(2,2|4)$ and
$\alhs(2,2|4)$ is the so called ``singleton{}''
$\mult_{[0,1,0][0,0]}^{1,0}$ associated to the ${\cal N}=4$ SYM
vector multiplet. Its HWS $|Z\rangle$, i.e.~the ground-state or
`vacuum', which is obviously different from the trivial
$\alPSU(2,2|4)$ invariant vacuum $|0\rangle$, is one of the
complex scalars, let us say $Z=\varphi^5 + i \varphi^6$. The other
(complex) components will be denoted by $X=\varphi^1 + i
\varphi^2$ and $Y=\varphi^3 + i \varphi^4$ in the following.
Showing that the singleton is an irreducible representation of
$\alPSU(2,2|4)$ is tantamount to showing that any elementary SYM
state can be found by acting on the Fock space vacuum $|Z\rangle$
with a sequence of superconformal generators chosen among
(\ref{supconfgen}). Looking at the singleton as an irrep of
$\alhs(2,2|4)$ one sees an important difference: the sequence of
superconformal generators \footnote{Without loss of generality we
may assume the length of the sequence to be odd; for an even
sequence we may append an element of the Cartan subalgebra,
e.g.~the dilatation generator.} is replaced by a single HS
generator and therefore any component $A$ in the singleton
multiplet can be reached in a single step $\mathcal{J}_{A\bar B}$
from any other one $B$. This can be shown by noticing that, since
the central charge $C$ commutes with all generators and
annihilates the vacuum, a non-trivial sequence in $({\cal
A}_1)^{2\ell+1}$ belongs to ${\cal A}_{2\ell+1}$. This difference,
irrelevant for one-letter states ($L=1$), will be crucial in
proving the irreducibility of
YT-pletons with respect to the HS algebra.%
\footnote{This property is also satisfied by the fundamental
representation of $SU(m)$. Our proof below
 reduces in this case to  the familiar statement that irreducible
 representations of $SU(m)$ are in one-to-one correspondence with
 Young tableaux made out of fundamentals.}

Let us now consider the tensor product of $L$ singletons. The
generators of $\alhs(2,2|4)$ are realized as diagonal
combinations: \[ {\cal J}_{2\ell+1}\equiv \sum_{s=1}^L\, {\cal
J}^{(s)}_{2\ell+1} \label{genl} \] with ${\cal J}_{2\ell+1}^{(s)}$
HS generators acting at the $s^{\rm th}$ site. The tensor product
of $L\geq 1$ singletons  is generically reducible not only under
$\alPSU(2,2|4)$ but also under $\hs$. This can be seen by noticing
that the HS generators (\ref{genl}), being completely symmetric,
commute with symmetrizations and antisymmetrizations of the
indices in the tensor product of singletons. In particular, the
tensor product decomposes into a sum of representations
characterized by Young tableaux $YT$ with $L$ boxes. A Young
tableaux is defined by distributing SYM letters among $L$ boxes
and acting on it with the operator ${\cal O}_{\rm YT}=A_{\rm YT}
S_{\rm YT}$ that first symmetrizes all letters in the same row and
then antisymmetrizes letters in the same column. This operator
clearly commutes with all generators of $\alhs(2,2|4)$, and
therefore different Young tableaux belong to different irreducible
components.

To prove irreducibility of $L$-pletons associated to a specific YT
with $L$ boxes under $\alhs(2,2|4)$, it is then enough to show
that any state in the $L$-pleton under consideration can be found
by acting on the relevant HWS with HS generators. Let us start by
considering states belonging to the totally symmetric tableau. The
simplest examples of such states are those with only \emph{one}
site different from the vacuum $Z$, i.e.~$AZ\ldots
Z+\mbox{symm.}$. Using the fact that any SYM letter $A$ can
reached from the HWS $Z$ using a single $\hs$ generator
$\mathcal{J}_{A\bar Z}$ we write the ``one impurity{}'' state as
$({\cal J}_{A\bar Z}Z)Z\ldots Z+{\rm symm.}$ This state can also
be written as ${\cal J}_{A\bar{Z}}(Z^L)$ and it is therefore a HS
descendant.
%Here $M=2\ell+1$ is understood, states with $2M=4\ell$ oscillators
%are accommodated in ${\cal J}_{2\ell+1} Z^L$ by inserting an extra
%element of the Cartan subalgebra.
%moved this comment up where it belongs.
 The next simplest class is given by states with
``two impurities{}'' $ABZ\ldots Z+{\rm symm.}$. Once again this
state can be written as ${\cal J}_{A\bar Z}{\cal J}_{B\bar
Z}(Z^L)$ up to the ``one impurity{}'' descendant $({\cal J}_{A\bar
Z}{\cal J}_{B\bar Z}Z)Z\ldots Z$ of the type already found.
Proceeding in this way the reader can easily convince him/herself
that all states in the completely symmetric tensor of $L$
singletons can be written as HS descendants of the vacuum $Z^L$.

The same arguments hold for generic tableaux. For example, besides
the descendants ${\cal J}_{A\bar{Z}} (Z^L)$ of $Z^L$ there are
$L-1$ ``one impurity{}'' multiplets of states associated to the
$L-1$ Young tableaux with $L-1$ boxes in the first row and a
single box in the second one\footnote{As we will momentarily see,
HS multiplets of this kind are absent for ${\cal N}=4$ SYM
theories with semisimple gauge group. At any rate, they are
instrumental to illustrate our point.}. The vacuum state of HS
multiplets associated to such tableaux can be taken to be
$Y_{(k)}\equiv Z^{k} Y Z^{L-k-1}-Y Z^{L-1}$ with $k=1,\ldots,L-1$.
Any state with one impurity $Z^{k} A Z^{L-k-1}-A Z^{L-1}$ with
$k=1,\ldots,L-1$ can be found by acting on $Y_{(k)}$ with the HS
generator ${\cal J}_{A\bar Y}$, where ${\cal J}_{A\bar Y}$ is the
HS generator that transforms $Y$ into $A$ (and annihilates $Z$).

Notice that the arguments rely heavily on the fact that any two
states in the singleton are related by a one-step action of a HS
generator. This is not the case for the ${\cal N}=4$ SCA, and
indeed the completely symmetric tensor product of $L$ singletons
is highly reducible with respect to $\alPSU(2,2|4)$, as we shall
see in the following.

\section{HS content of ${\cal N}=4$ SYM}
\label{smult}

The on-shell field content of the singleton representation of
$\alPSU(2,2|4)$ is encoded in the partition function
\begin{eqnarray}
\Yboxdim4pt {\cal Z}_{\yng(1)}\,(t,y_i) &=& \sum_{s=0}^\infty \,
\left[ t^{1+s}\,\chi_{[\frac{s}{2},\frac{s}{2}]}\,\chi_{[010]}~+
t^{2+s}\,\chi_{[\frac{s+2}{2},\frac{s}{2}]}\,\chi_{[000]}~+
t^{2+s}\,\chi_{[\frac{s}{2},\frac{s+2}{2}]}\,\chi_{[000]}~+
\right.\nn\\
&& \left.-
t^{\frac{3+s}2}\,\chi_{[\frac{s+1}{2},\frac{s}{2}]}\,\chi_{[001]}~-
t^{\frac{3+s}2}\,\chi_{[\frac{s}{2},\frac{s+1}{2}]}\,\chi_{[100]}
\right]\;, \label{sing} \end{eqnarray}
with the different terms corresponding to the six real scalars
$\varphi^i$, the field strengths $F^{\pm}_{\mu\nu}$ and the
fermions $\lambda_A^\alpha$, $\bar\lambda^A_{\dot{\alpha}}$,
respectively, together with their derivatives. Here $t$ keeps
track of the bare conformal dimension $\Delta$.
$\chi_{[j,\jb]}\chi_{[q_1,p,q_2]}(y_i)$ denotes the character
polynomial of the $\mathfrak{so}(4)\times \mathfrak{so}(6)$
representation $[j,\jb] [q_1,p,q_2]$\footnote{Characters of UIR's
of superconformal algebras have been recently derived in
\cite{Dobrev2}.}. In particular, focusing only on the scaling
dimensions $\Delta$ and performing explicitly the sum over $s$,
one finds the one-letter partition function\footnote{At $y_i=1$
one has by definition $\chi_{[q_1,p,q_2]}={\rm dim}[q_1,p,q_2] =
(q_1+1)(p+1)(q_2+1)(p+q_1+2)(p+q_2+2) (p+q_1 + q_2 + 3)/12$ and $
\chi_{[j,\jb]}={\rm dim}[j,\jb]=(2j+1)(2\jb+1)$.}
\begin{eqnarray}
\Yboxdim4pt {\cal Z}_{\yng(1)}\,(t,y_i)|_{y_i=1} &=&  \frac{ 2\,
t\, (3+ t^{\frac{1}{ 2}})}{ (1+t^{\frac{1}{2}})^3} \;.
\end{eqnarray}
As explained above, the singleton turns out to be the
``fundamental representation{}'' of $\alhs(2,2|4)$ as well.
Moreover, we have argued that representations of $\alhs(2,2|4)$
are built in terms of tensor products of singletons properly
decomposed according to irreducible representations of the
permutation group. These are associated to Young Tableaux built
from $\Yboxdim6pt \yng(1)\equiv \Yboxdim4pt{\cal
Z}_{\yng(1)}\,(t)$. The spectrum of single-trace operators in
${\cal N}=4$ SYM theory with $SU(N)$ gauge group is given by all
possible {\em cyclic} words built from letters chosen
from~$\Yboxdim6pt{\cal Z}_{\yng(1)}$. It can be computed using
Polya theory~\cite{Polya}, which gives the generating
function~\cite{Sundborg:1999ue,Sundborg:2000wp,Haggi-Mani:2000ru,Polyakov:2001af,Bianchi:2003wx}
\[
\Yboxdim4pt {\cal Z}({u},t,y_i)= \sum_{n\geq 2}\, u^n\,{\cal Z}_n
(t,y_i)= \sum_{n\geq 2,d|n}\,u^n\,\frac{\varphi(d)}{n}\,{\cal
Z}_{\yng(1)}\,(t^d,y_i^d)^{\frac{n}{d}} \;, \label{polyah2}
\]
for cyclic words. Here $u$ keeps track of the length $L$, i.e.~the
number of letters / partons. The sum runs over all integers $n>2$
and their divisors $d$, and $\varphi(d)$ is Euler's totient
function, that equals the number of integers smaller than and
relatively prime to $d$. For later convenience, we have introduced
the notation ${\cal Z}_n(t,y_i)$ to denote the restriction to
cyclic words made out of $n$-letters. The partition function
(\ref{polyah2}) accounts for SYM composite operators and all their
derivatives, i.e.~their $\alSO(4,2)/(\alSO(4)\!\times\!\alSO(2))$
descendants. $\alSO(4,2)$ primaries can instead be read off from
$\widehat{\cal Z}({u},t,y_i)$, defined from ${\cal Z}({u},t,y_i)$
by removing total derivatives:
\begin{eqnarray}
\widehat{\cal Z}({u},t,y_i) &\equiv& {\cal Z}({u},t,y_i)\,
\left(1-t\, \chi_{[\frac12\frac12]}+t^2\,(\chi_{[10]}+\chi_{[01]})
-t^3\,\chi_{[\frac12\frac12]}+t^4\right) \;. \label{totder}
\end{eqnarray}
We note that $\Yboxdim4pt {\cal Z}_{\yng(1)}\,({u}^d,t^d,y_i^d)$
denotes the alternating sum over length-$d$ Young tableaux of the
hook type:
\[
\Yboxdim4pt {\cal Z}_{\yng(1)}\,(t^d) ~=~ {\cal
Z}_{\yng(5)\cdot\cdot\yng(2)}(t)~-~ {\cal
Z}_{\yng(4,1)^{\cdot\cdot\yng(2)}}(t)~+~ {\cal
Z}_{\yng(3,1,1)^{^{\cdot\cdot\yng(2)}}}(t)~-~ {\cal
Z}_{\yng(2,1,1,1)^{^{^{\cdot\cdot\yng(2)}}}}(t) ~+~ \ldots \;.
\label{hook}
\]
 Plugging this expansion into
(\ref{polyah2}), we find for the first few cases:
\begin{eqnarray}
\Yboxdim4pt {\cal Z}_2 &=& \Yboxdim4pt
{\cal Z}_{\yng(2)}\;,\nn\\[1ex]
{\cal Z}_3 &=& \Yboxdim4pt {\cal Z}_{\yng(3)}+{\cal
Z}_{\yng(1,1,1)}
\;,\nn\\[1ex]
{\cal Z}_4 &=& \Yboxdim4pt {\cal Z}_{\yng(4)}+ {\cal
Z}_{\yng(2,1,1)}+{\cal Z}_{\yng(2,2)}
\;,\nn\\[1ex]
{\cal Z}_5 &=& \Yboxdim4pt {\cal Z}_{\yng(5)}+{\cal
Z}_{\yng(3,2)}+ 2\, {\cal Z}_{\yng(3,1,1)}+ {\cal
Z}_{\yng(2,2,1)}+ {\cal Z}_{\yng(1,1,1,1,1)}\;, \qquad\mbox{etc.}
\label{hsD}
\end{eqnarray}

Notice that only a subset of YT, those compatible with cyclicity
of the trace, enters in (\ref{hsD}). In particular, HS multiplets
associated to the tableaux $\Yboxdim6pt\yng(1,1)$,
$\Yboxdim6pt\yng(2,1)$, two out of the three of type
$\Yboxdim6pt\yng(2,1,1)$, and so on, are projected out. The
content of the various components in (\ref{hsD}) can be derived
from the formulae:
 \begin{eqnarray}
\Yboxdim4pt {\cal Z}_{\yng(2)}&=&\Yboxdim4pt \frac{1}{
2!}\left[{\cal Z}_{\yng(1)}\,(t)^2+ {\cal
Z}_{\yng(1)}\,(t^2)\right]\nn\\
\Yboxdim4pt {\cal Z}_{\yng(3)}&=&\Yboxdim4pt \frac{1}{
3!}\left[{\cal Z}_{\yng(1)}\,(t)^3+ 3\,{\cal
Z}_{\yng(1)}\,(t^2){\cal Z}_{\yng(1)}\,(t)
+2\,{\cal Z}_{\yng(1)}\,(t^3)\right]\nn\\
\Yboxdim4pt {\cal Z}_{\yng(1,1,1)} &=&\Yboxdim4pt \frac{1}{
3!}\left[{\cal Z}_{\yng(1)}\,(t)^3- 3\,{\cal
Z}_{\yng(1)}\,(t^2){\cal Z}_{\yng(1)}\,(t)+2\,{\cal
Z}_{\yng(1)}\,(t^3)\right]\nn\\
\Yboxdim4pt {\cal Z}_{\yng(4)}&=& \Yboxdim4pt \frac{1}{
4!}\left[{\cal Z}_{\yng(1)}\,(t)^4+ 6\,{\cal
Z}_{\yng(1)}\,(t^2){\cal Z}_{\yng(1)}\,(t)^2+3\,{\cal
Z}_{\yng(1)}\,(t^2)^2+ 8\,{\cal Z}_{\yng(1)}\,(t^3){\cal
Z}_{\yng(1)}\,(t)+6\,{\cal
Z}_{\yng(1)}\,(t^4)\right]\nn\\
\Yboxdim4pt {\cal Z}_{\yng(2,2)}&=&\Yboxdim4pt \frac{1}{
4!}\left[2\, {\cal Z}_{\yng(1)}\,(t)^4+ 6\,{\cal
Z}_{\yng(1)}\,(t^2)^2-8\,{\cal Z}_{\yng(1)}\,(t^3)
\,{\cal Z}_{\yng(1)}(t)\right]\nn\\
\Yboxdim4pt {\cal Z}_{\yng(2,1,1)}&=&\Yboxdim4pt
\frac{1}{4!}\left[3\, {\cal Z}_{\yng(1)}\,(t)^4-6\,{\cal
Z}_{\yng(1)}\,(t^2) {\cal Z}_{\yng(1)}\,(t)^2- 3\,{\cal
Z}_{\yng(1)}\,(t^2)^2+6\,{\cal Z}_{\yng(1)}\,(t^4)\right]
\label{ds} \;.
\end{eqnarray}
Formulae (\ref{ds}) can be explicitly verified with the use of
(\ref{hook}).

Under the superconformal group $\alPSU(2,2|4)$, the HS multiplet
${\cal Z}_{YT}$, associated to a given Young tableau $YT$ with $L$
boxes, decomposes into an infinite sums of multiplets. The HWS's
can be found by computing ${\cal Z}_{YT}$ and eliminating the
superconformal descendants by passing ${\cal Z}_{YT}$ through a
sort of Erathostenes' (super) sieve~\cite{Bianchi:2003wx}. This
will be the subject of the next subsection. Here we just state the
results for $L=2,3$. The complete list of $\alPSU(2,2|4)$
multiplets appearing in the decomposition of the first few HS
multiplets with $L=2,3$ letters is collected in
table~\ref{tnotations}, see Appendix~A for the notation of
$\alPSU(2,2|4)$ multiplets.
\begin{table}\centering
$\begin{array}[h]{|c|c|c|c|}\hline L  & {\rm name} &
 \mult^{\Delta,B}_{[j,\jb][q_1,p, q_2]} & {\rm
sector}
\\\hline\hline
2& \mult_{0} & \mult^{2,0}_{[0^\dagger,0^\dagger][0,2,0]} &
\alSL(2)_{j=\mis\frac{1}{2}}\\\hline 2 & \mult_{n} &
\mult^{n,0}_{[\frac{1}{2} n\mis1^\ast,\frac{1}{2}
n\mis1^\ast][0,0,0]}
 & \alSL(2)_{j=\mis\frac{1}{2}}\\\hline
3 &  \mult_{0,0} & \mult^{3,0}_{[0^\dagger,0^\dagger][0,3,0]}
 & \alSL(2)_{j=\mis\frac{1}{2}}\\\hline
3 & \mult_{0,n}
&\mult^{n+1,0}_{[\frac{1}{2}n-1^\ast,\frac{1}{2}n-1^\ast][0,1,0]}
 &  \alSL(2)_{j=\mis\frac{1}{2}}\\\hline
3  & \mult_{1,n}&
\mult^{n+\frac{5}{2},+\frac{1}{2}}_{[n/2^\ast,n/2-1/2^\ast][0,0,1]}
 & \alSL(2)_{j=\mis1}\\\hline
3 & \mult_{-1,n}&
\mult^{n+\frac{5}{2},-\frac{1}{2}}_{[n/2-1/2^\ast,n/2^\ast][1,0,0]}
&  \alSL(2)_{j=\mis1}\\\hline 3 & \mult_{m\geq +2,n}&
\mult^{n+2m,1}_{[\frac{1}{2}n+m-1^\ast,\frac{1}{2}n][0,0,0]} &
\alSU(1,2)\\\hline 3&\mult_{m\leq -2,n}&
\mult^{n+2m,-1}_{[\frac{1}{2}n,\frac{1}{2}n+m-1^\ast][0,0,0]}
&\alSU(1,2)\\
\hline
\end{array}$
\caption{$\alPSU(2,2|4)$ multiplets with $L\leq 3$.}
\label{tnotations}
\end{table}
The decompositions of the corresponding HS multiplets reads:
\< \Yboxdim4pt {\cal Z}_{\yng(2)}\eq \sum_{n=0}^\infty \mult_{2n}
\;, \qquad \Yboxdim4pt {\cal Z}_{\yng(1,1)}=
\sum_{n=0}^\infty\mult_{2n+1} \;, \nln
\Yboxdim4pt {\cal Z}_{\yng(3)} \eq
\sum_{k=-\infty}^\infty\sum_{n=0}^\infty c_n\left[\mult_{2k,n}
+\mult_{2k+1,n+3} \right] \;,\nln
\Yboxdim4pt {\cal Z}_{\yng(2,1)} \eq
\sum_{k=-\infty}^\infty\sum_{n=0}^\infty d_n\left[\mult_{2k,n+1}
+\mult_{2k+1,n+1} \right] \;,\nln
\Yboxdim4pt {\cal Z}_{\yng(1,1,1)}\eq
\sum_{k=-\infty}^\infty\sum_{n=0}^\infty c_n\left[\mult_{2k,n+3}
+\mult_{2k+1,n} \right] \;.\label{123} \> The coefficients
$c_n\equiv 1+[n/6]-\delta_{n,1~{\rm mod}~6}$ and $d_n\equiv
1+[n/3]$ with $[m]$ the integral part of $m$, are the
multiplicities of $\alPSU(2,2|4)$ multiplets inside
$\alhs(2,2|4)$. More precisely $c_n,d_n$ count the number of ways
one can distribute derivatives (HS descendants) between the boxes
in the tableaux. These multiplicities will be computed in the next
section, cf.~\eqref{cnco} below. For convenience of the reader we
display the translation of these formulae into $\alPSU(2,2|4)$
notation $\mult^{\Delta,B}_{[j,\jb][q_1,p,q_2]}$ in
Appendix~\ref{a123}.

The multiplets with $n=0$ or $m=0,\pm 1$ in table~\ref{tnotations}
are special: $n=0$ corresponds to the $\ft12$-BPS series, dual to
${\cal N}=8$ gauged supergravity and its KK recurrences, $m=0,\pm
1,n\geq 1$ to semishort-semishort multiplets. Finally for $m\geq
2$ one finds multiplets satisfying a bound of type long-semishort.

The `symmetric doubleton' $\Yboxdim4pt{\cal Z}_{\yng(2)}$
 contains the multiplets of conserved HS currents $\mult_{2n}$.
 The `antisymmetric doubleton' $\Yboxdim4pt{\cal Z}_{\yng(1,1)}$  is
ruled out by cyclicity of the trace, cf.~\eqref{hsD}.  The
`symmetric tripleton' $\Yboxdim4pt{\cal Z}_{\yng(3)}$
(corresponding to the cubic Casimir $d_{abc}$) contains the first
KK recurrences of twist 2 semishort multiplets, the still
semishort-semishort series $\mult_{\pm 1,n}$ starting with
fermionic primaries and long-semishort multiplets.
 The `antisymmetric tripleton' $\Yboxdim4pt{\cal Z}_{\yng(1,1,1)}$
 (corresponding to the structure constants $f_{abc}$)
on the other hand contains the Goldstone multiplets that join to
multiplets with twist 2 to form long multiplets when the HS
symmetry is broken. In addition, fermionic semishort-semishort
multiplets and long-semishort multiplets also appear.

\section{Partition function of semishort superprimaries}
\label{ssemishort}

In this section, we focus on the particularly interesting class of
SYM operators sitting in BPS and semishort multiplets of the
superconformal algebra $\alPSU(2,2|4)$ and derive multiplicity
formulae for their superprimaries. Semishort and BPS multiplets
are special in that their components encompass all generalized
`massless' states and their superpartners. By this we mean SYM
operators whose dimensions saturate unitary bounds and whose
holographic duals would thus be massless in a manifestly
$SO(10,2)$ symmetric description in the bulk
\cite{Bars:2002pe,Beisert:2003te,Brink:2000ag}. Not unexpectedly,
we will find that general formulae drastically simplify for these
operators. When interactions are turned on ($g_{YM}\neq 0$),
i.e.~departing from the HS enhancement point, only truly 1/2 BPS
multiplets remain `massless' in the above generalized sense. All
semishort multiplets participate in the `Grande Bouffe', whereby
they `eat' the relevant Goldstone / St\"uckelberg multiplets and
become massive. The resulting long multiplets acquire anomalous
dimensions and, in principle, mix with one another compatibly with
their quantum numbers.

A generic long $\alPSU(2,2|4)$ multiplet will be denoted as
$\mult_{[q_1,p,q_2]{[j,\jb]}}^{\Delta,B}$ by means of the Dynkin
labels of its HWS with respect to the compact bosonic subalgebra
$\mathfrak{su}(4)\times \mathfrak{su}(2)^2\times
\mathfrak{u}(1)_{\Delta}$ and the `external' $\mathfrak{u}(1)_B$
hypercharge. More precisely, $[q_1,p,q_2]$ are Dynkin labels of
$\mathfrak{su}(4)$ while $[j,\jb]$ denote the spins under
$\mathfrak{su}(2)\times \mathfrak{su}(2)$. At particular values of
$\Delta$, the long multiplet
$\mult_{[q_1,p,q_2]{[j,\jb]}}^{\Delta,B}$ may split into
semi-short or BPS multiplets, cf.~Appendix~A for details.

For the following, it is convenient to split superoscillator
indices with respect to the $\mathfrak{su}(2)_a\times
\mathfrak{su}(2)_b\times \mathfrak{su}(2)_c\times
\mathfrak{su}(2)_d$ subalgebra inside $\mathfrak{su}(2,2)\times
\mathfrak{su}(4)$, which yields
$y_a=(a_\alpha,-b^\dagger_{\dot{\alpha}})$,
$\bar{y}^a=(a_\alpha^\dagger,b_{\dot{\alpha}})$,
$\theta_A=(c_r,d^\dagger_{\dot{r}})$,
$\bar{\theta}^A=(c_r^\dagger,d_{\dot{r}})$, with indices
$\alpha,\dot{\alpha},r,\dot{r}$ taking values $1,2$ . In this
notation, the basic representation,  the singleton is denoted
as~$\mult_{[0,1,0][0,0]}^{1,0}$. Its HWS $|Z\rangle$, i.e.~the
ground-state or `vacuum', is chosen to be the scalar component
$Z=\varphi^5 + i \varphi^6$ that satisfies  \[
{a}_{\alpha}|Z\rangle ={b}_{\dot\alpha}|Z \rangle
={c}_{r}|Z\rangle = { d}_{\dot{r}}|Z \rangle =0 \;. \] and is thus
invariant under the non-semisimple superalgebra that combines
$\mathfrak{iso}(4)_{ab}\times
\mathfrak{iso}(4)_{cd}\times\mathfrak{u}(1)_{\Delta - J}\times
\mathfrak{u}(1)_{C}$ with 24 supercharges (16 $S$'s and 8 $Q$'s).
Clearly $|Z\rangle$ cannot be obtained from the
$\mathfrak{su}(2,2|4)$ invariant trivial, but still physical,
vacuum $|0\rangle$, associated to the identity operator, through
the action of a finite number of oscillators.

Physical states in the singleton representation are given by all
possible excitations $(a^\dagger)^{n_a}
(b^\dagger)^{n_b}(c^\dagger)^{n_c} (d^\dagger)^{n_d}|Z\rangle$
satisfying the zero central charge condition \[ {n}_a- {n}_b +
{n}_c - {n}_d  = 0 \;. \label{zoc}
 \]
One can easily check that all elementary fields of ${\cal N}=4$
SYM and their derivatives can be represented in this way. The six
scalars $\varphi^i$ are given by the vacuum together with the
excitations $c^\dagger_r d^{\dagger}_{\dot{r}}$, $c^\dagger_1
c^\dagger_2 d^{\dagger}_{1}d^{\dagger}_{2}$. The left-handed
gaugini $\lambda_A^\alpha$ by the excitations $a^{\dagger}_\alpha
d^{\dagger}_{\dot{r}}$ and $a^{\dagger}_\alpha c^{\dagger}_r
d^{\dagger}_{1}d^{\dagger}_{2}$. The right-handed gaugini
$\bar\lambda^A_{\dot\alpha}$ by $b_{\dot\alpha}^\dagger
c^{\dagger}_r$ and $b_{\dot\alpha}^\dagger
d^{\dagger}_{\dot{r}}c^{\dagger}_1 c^{\dagger}_2$. The field
strengths $F_{\mu\nu}^\pm$ by $a^\dagger_\alpha a_\beta^\dagger
d^{\dagger}_{1}d^{\dagger}_{2}$, $b^\dagger_{\dot\alpha}
b_{\dot\beta}^\dagger c^\dagger_1 c^\dagger_2$. Finally,
space-time derivatives are given by the action of
$P_{\alpha\dot\alpha}= a^\dagger_\alpha b_{\dot\alpha}^\dagger$.

For the tensor product of $L$ singletons, oscillators
$a_\alpha^{(s)}$,$b_{\dot{\alpha}}^{(s)}$,$c_r^{(s)}$,
$d_{\dot{r}}^{(s)}$ are to be thought as length $L$ vectors with
components acting at each of the $L$ sites and trivial
(anti-)commutation relations between oscillators acting on
different sites. The vacuum $Z^L$ is the tensor product of $L$
copies of the singleton vacuum $|Z\rangle^L$. The Dynkin labels
$[j,\jb][q_1,p,q_2]^{\Delta,B}$ of a length $L$ SYM state made out
of $n_a$, $n_b$, $n_c$ and $n_d$ oscillators follow from the
relations
\begin{eqnarray} &&\Delta = L+ \ft12 n_a+\ft12 n_b\;,\quad B = \ft12
n_{d}-\ft12 n_{c} =\vert_{_{C=0}} \ft12
n_{a}-\ft12 n_{b} \;,\nn\\[.5ex]
&&\left[j,\jb\,\right] =  \left[\ft12 (n_{a_1}\!-\!n_{a_2}), \ft12
(n_{b_1}\!-\!n_{b_2})  \right] \;,
\nn\\[.5ex]
&&\left[q_1,p,q_2\,\right] = \left[n_{c_2}-n_{c_1},
L-n_{c_2}-n_{d_1} ,n_{d_1}-n_{d_2} \right] \;, \label{DszI}
\end{eqnarray}
with $n_a,n_b,n_c,n_d$, the total number of oscillators of a given
type. In addition the zero central charge condition (\ref{zoc}),
i.e.~$C_{(s)}=0$, is imposed  at each site $s$.

\subsection{Restricted semishort multiplets }

The oscillator numbers $n_a$, $n_b$, $n_c$, $n_d$ in (\ref{DszI})
are required to be positive, since $|Z\rangle_L$ is annihilated by
all raising operators. This simple condition imposes non-trivial
bounds on the allowed $\psu$ charges in the SYM spectrum. For
example $n_{a_2} + n_{c_1}\geq 0$ and $n_{b_2} + n_{d_2}\geq 0$
together with \eqref{zoc} imply the lower bounds
\begin{eqnarray}
\Delta~\ge~ 2j+\ft32 q_1+p+\ft12 q_2\;,\qquad \Delta~\ge~
2\bar{\jmath}+\ft12 q_1+p+\ft32 q_2\;, \label{BPS}
\end{eqnarray}
for the conformal dimension of any state (not only HWS's!). In
this section we will focus on states that simultaneously saturate
the two bounds~\eqref{BPS}, or equivalently satisfy the
intersection condition
\begin{eqnarray}
\Delta&=& p+q_1+q_2+j+\jb\;.\label{rbound}
\end{eqnarray}
This kind of states are only present in BPS and semishort
multiplets. This can be seen by noting that the field content of
any multiplet is generated by acting on the HWS with (a subset of)
the 16 supercharges ${\cal Q}^A{}_\alpha$, $\bar{{\cal
Q}}_{A\dot{\alpha}}$, cf.~Appendix~A. The only supersymmetry
charges among (\ref{qs}) whose weights violate the bounds
(\ref{BPS}) are $Q_1^+,\bar{Q}_4^+$ and they do so by exactly one
unit. Therefore a state satisfying (\ref{rbound}) should belong to
a multiplet whose HWS has a conformal dimension that exceeds
(\ref{rbound}) by at most two units, i.e.~$\Delta\leq
2+p+q_1+q_2+j+\jb$. This happens only for BPS or semishort
multiplets. Indeed, the state under consideration could either be
the HWS of a BPS multiplet that satisfies (\ref{rbound}) and is
annihilated by $Q_1^+,\bar{Q}_4^+$ or the level two
superdescendant, \[\label{QQ14}|\Psi_{2}\rangle = Q_1^+
\bar{Q}_4^+ |\Psi_{0}\rangle,\] in a semishort multiplet whose HWS
$|\Psi_{0}\rangle$ has $\Delta_{0}=2+p+q_1+q_2+j+\jb$.

We will conveniently use states satisfying \eqref{rbound} as
representatives of semishort and BPS multiplets. In terms of
oscillators, this bound amounts to restricting attention to states
for which
\[
n_{a_2}=n_{b_2}=n_{c_1}=n_{d_2}=0 \label{ns0} \;.
\]
For simplicity, in the following, we denote the surviving
oscillators $(a_1,b_1,c_2,d_1)$ simply by $(a,b,c,d)$. From
\eqref{DszI} it follows that a SYM state with Dynkin labels
$[j,\jb][q_1,p,q_2]^{\Delta,B}$ satisfying~\eqref{rbound} carries
 \begin{eqnarray}
\Delta= L+j+\jb\;, \qquad L= p+q_1+ q_2 \qquad B
=\ft12(q_2\!-\!q_1) \;, \label{lsu2}\end{eqnarray}
and will be represented by the oscillator monomial
\begin{eqnarray}
[j,\jb][q_1,p,q_2]^{\Delta,B} &\equiv&  a^{2 j} \,b^{2 \jb}
\,c^{q_1} \,d^{q_2} \, y^{p+q_1+q_2} \;. \label{weight}
\end{eqnarray}
The letters $a,b,c,d$ here have a two-fold meaning. On the one
hand they keep track of the quantum numbers $q_1,q_2,j,\jb$, on
the other hand they describe how a given state is made out of
oscillators $a$,$b$,$c$,$d$. Finally, the auxiliary variable $y$
keeps track of $p$. Notice that for states satisfying
(\ref{rbound}), $p$ is related to the number of letters
$L=p+q_1+q_2$ via (\ref{lsu2}), and therefore powers of $y$
simultaneously count the number of letters, previously counted
by~${u}$.

On these states, the residual superconformal symmetry is
$\alSU(1,1|2)\subset\alPSU(2,2|4)$. The $\alSU(1,1|2)$ raising
operators among \eqref{supconfgen} are
 \[
 Q_2^+={\frac{a}{c}}\;,
\quad \bar{Q}_2^+= b c \;, \quad Q_3^+=a d\;, \quad
 \bar{Q}_3^+={\frac{b}{d}}\;,
 \qquad P=a b\;,
 \quad
 J=c d\;,
 \label{genr} \]
preserving the bound~\eqref{rbound}. Positive and negative powers
in these expressions are associated to creation and annihilation
operators respectively, e.g.~${\frac{a}{c}}\equiv a_1^\dagger c_2,
{\frac{b}{d}}\equiv b_1^\dagger d_1$, and so on. It is then
convenient to consider for BPS and semishort multiplets instead of
the full character polynomials of $\alPSU(2,2|4)$ and its bosonic
subgroup $\alSO(4)\times \alSO(6)$, the restriction to states
satisfying~\eqref{rbound}, giving rise to character polynomials of
$\alSU(1,1|2)$ and its bosonic subgroup $\alSL(2)\times \alSU(2)$,
respectively. We denote these as ${\cal V}^{\rm
rst,\Delta}_{[j,\jb][q_1,p,q_2]}$ and ${\chi}^{\rm
rst}_{[j,\jb][k,p,q]}$, respectively. Discarding from now on
$\alSL(2)$ descendants, i.e.~total derivatives generated
by~$P=ab$, the character polynomial exclusively generated by the
$\alSU(2)$ raising operator $J=cd$ reads
\begin{eqnarray}
{\chi}^{\rm rst}_{[j,\jb][q_1,p,q_2]} &=& a^{2j} b^{2\jb} c^{q_1}
d^{q_2} y^{p+q_1+q_2}\;
 \frac{1-(c d)^{p+1}}{1-cd} \;.
\label{su2}
\end{eqnarray}
 As discussed above, the
restricted character polynomials  ${\cal V}^{\rm
rst,{\Delta}}_{[j,\jb][q_1,p,q_2]}$ is non-trivial only for BPS
and semishort multiplets. For semishort multiplets one finds
\begin{eqnarray}
{\cal V}^{\rm rst,{\Delta}}_{[j^*,\jb^*][q_1,p,q_2]} &=&
{\chi}^{\rm
rst}_{[q_1+1,p,q_2+1](j+\frac12,\bar{\jmath}+\frac12)}\, T_{\rm
short} ~=~ y^2 abcd\,{\chi}^{\rm
rst}_{[q_1,p,q_2](j,\bar{\jmath})}\, T_{\rm short}
 \;,
\label{semi}
\end{eqnarray}
with
\[
T_{\rm short}=(1-ad)(1-bc)(1-{\frac{a}{c}})(1-{\frac{b}{d}}) \;,
\label{kon}
\]
generated by the four $\alPSU(1,1|2)$ supercharges (\ref{genr}).
The factor $y^2 abcd$ takes care of the highest weight states of
the restricted semishort multiplets, cf.~\eqref{QQ14}, and maps
$\alPSU(1,1|2)$ primaries to semishort ${\cal N}=4$ superconformal
primaries. The number of states inside the multiplet~\eqref{semi}
is given by $2^4$ times the restricted dimension of the highest
weight state, i.e.~$2^4\,(p+1)$. The nice factorized form
(\ref{semi}) of the restricted semishort multiplet is to be
contrasted with the more involved multiplicity formulae for
semishort multiplets in $\alPSU(2,2|4)$. We will make use of this
restriction as a powerful simplifying tool in our analysis. The
simplest generic multiplet of type~\eqref{semi} is the restriction
of the short Konishi multiplet
\[
 {\cal V}^{\rm rst,2}_{[0,0][0,0,0]}=y^2 abcd \,T_{\rm short}
 \;,
\]
with total dimension $2^4$. Notice that the state $y^2 abcd$,
corresponding to the weight $[101][{\frac12},{\frac12}]$, is the
highest component of the Konishi current with $\Delta_0=3$ in the
${\bf 15}=[1,0,1]$ of $SU(4)$ that is a singlet ($p=0$) of
$\alSU(2)\subset \alPSU(1,1|2)$.

The factorized formula (\ref{semi}) also holds for the
$\frac14$-BPS multiplets which are counted according to
\eqref{bps} below. In contrast, the restricted character
polynomial corresponding to the $\frac12$-BPS multiplet ${\cal
V}^{{\rm rst},n}_{{[0^\dagger ,0^\dagger]}[0n0]}$ is generated by
$J$ and the supersymmetry charges
$Q_3^+,\bar{Q}_2^+$.\footnote{The full $\frac12$-BPS multiplet is
generated by $Q_{3,4}^{\pm},\bar{Q}_{1,2}^{\pm}$ supersymmetries
and $\alSU(4)\times \alSO(4)$ charges, cf.~Appendix~A.} With
(\ref{genr}) one finds:
  \begin{eqnarray}
 {\cal V}^{{\rm rst},n}_{{[0^\dagger,0^\dagger]}[0,n,0]} &=&
 {\chi}^{\rm rst}_{[0,0][0,n,0]}
 -{\chi}^{\rm rst}_{[\frac12,0][0,n,1]}
 - {\chi}^{\rm rst}_{[0,\frac12][1,n,0]}
 + {\chi}^{\rm rst}_{[\frac12,\frac12][1,n,1]}
 \nn\\[1ex]
&=& y^n \,{\frac{(1-ad)(1-bc)-(cd)^{n}(a- c)(b-d)}{(1-cd)}} \;.
\end{eqnarray}

\subsection{The semishort primary sieve}

Here we derive multiplicity formulae for semishort-semishort
$\alPSU(2,2|4)$ multiplets in ${\cal N}=4$ SYM theory. According
to (\ref{ns0}) the spectrum of single-letter SYM words saturating
the bound (\ref{rbound}) consists of all possible excitations
satisfying (\ref{ns0}). The multiplicities of these states can be
derived via Polya theory. The basic ingredient is the one-letter
partition function:
\[
{\cal Z}^{\rm rst}_1=y \,\frac{1+cd-ad-bc}{1-ab} \;,
\]
obtained from \eqref{sing} upon restriction. The four terms in the
numerators corresponds to the elementary SYM fields saturating the
bound (two scalars and two fermionic components) while the
expansion of the denominator generates their derivatives. The
restricted partition function is given by Polya's formula
\eqref{polyah2}:
\begin{eqnarray}
{\cal Z}^{\rm rst}_n &=& y^n(1-a b)\, \sum_{d|n}
\frac{\varphi(d)}{n}\left[ \frac{1+ (cd)^d-(ad)^d-(bc)^d}
{1-(ab)^d }\right]^{n/d} \label{zn} \;,
\end{eqnarray}
The factor $(1-a b)$ removes total derivatives, in much the same
way as in~\eqref{totder}. The restricted polynomial (\ref{zn})
contains only contributions coming from $\ft12$-BPS and semishort
multiplets. This can be checked by noticing that once the BPS
series $\sum_n {\cal V}^{{\rm
rst},n}_{[0^\dagger,0^\dagger][0,n,0]}$ is subtracted, the
spectrum organizes into multiplets of the type (\ref{semi}).
Specifically, the difference $({\cal Z}^{\rm rst}_n- {\cal
V}^{{\rm rst},n}_{[0^\dagger,0^\dagger][0,n,0]})$ vanishes at the
four zeros of \eqref{kon}
 \begin{eqnarray}
({\cal Z}^{\rm rst}_n- {\cal V}^{{\rm
rst},n}_{[0^\dagger,0^\dagger][0,n,0]})\Big|_{a=c,\frac1{d}}=
({\cal Z}^{\rm rst}_n- {\cal V}^{{\rm
rst},n}_{[0^\dagger,0^\dagger][0,n,0]})\Big|_{b=d,\frac1{c}}=0 \;,
\end{eqnarray}
as follows from the remarkable identity $\sum_{n|d}\,
\varphi(d)=n$. Semishort primaries can then be isolated by
factoring out $T_{\rm short}$. More precisely,
\begin{eqnarray}
{\cal Z}^{\rm short}_{n, \rm suprim} &\equiv&
%\frac{1}{y^2\,abcd}\, T^{-1}_{\rm short}
(y^2\,abcd\, T_{\rm short})^{-1} \,\left( {\cal Z}^{\rm
rst}_n-{\cal V}^{\rm rst}_{[0^\dagger,0^\dagger][0,n,0]}
\right)_{\rm HW}+\frac{y^{n-2}}{a^2 b^2} \;, \label{Zrstres}
\end{eqnarray}
is a regular rational function describing the character polynomial
of superprimaries sitting in
 semishort and BPS multiplets in the $n$-letter spectrum of SYM states.
$(y^2\,abcd\, T_{\rm short})^{-1}$ disposes of supersymmetry
descendants according to (\ref{semi})\footnote{In particular the
factor $y^2\,abcd$ map $\alSU(1,1|2)$ HWS to superconformal
primaries via (\ref{QQ14}).}. The subscript HW denotes the
reduction to $\alSU(2)$ highest weight states given by dividing
out the $\alSU(2)$ multiplets~\eqref{su2}. This can be done by
counting states according to the rule
\begin{eqnarray}
y^{p+q_1+q_2} c^{q_1} d^{q_2} \rightarrow \left\{
\begin{array}{ll}  y^{p+q_1+q_2}
c^{q_1} d^{q_2} & p\ge 0 \\
 -y^{p+q_1+q_2} c^{q_1+p+1} d^{q_2+p+1}\,  \quad & p< 0
\end{array} \;\right.
\;,
\end{eqnarray}
isolating $\alSU(2)$ HWS's. Alternatively the same result is found
by multiplying ${\cal Z}^{\rm rst}_{n, \rm suprim}$ by $(1- c d )$
and then deleting all bosons (fermions) coming with negative
(positive) multiplicities. The term ${y^{n-2}}/{a^2 b^2}$ in
\eqref{Zrstres} accounts for $\ft12$-BPS primaries with weights
${[-1,-1]}{[0,n-2,0]}={[00][0n0]}$ according to (\ref{bpsnot}).
% Finally the factor $(Q_1^+\bar{Q}^+)\equiv {a b c d\,y^2}$ with weight
%$[\ft12,\ft12][101]$,
% maps $\alPSU(1,1|2)$ primaries to ${\cal N}=4$ superconformal primaries.
 Notice that here powers of $y$ are no longer related to the number
  of letters (powers of $\ell$) since semishort primaries do not belong to the $\alPSU(1,1|2)$
sector.

For the lowest values of $L$, the above procedure yields
\begin{eqnarray}
\Yboxdim4pt  {\cal Z}^{\rm short}_{\yng(2), \rm suprim} &=&
\frac{1}{a^2 b^2(1-a^2b^2)} \;,
\label{Zresult}\\
\Yboxdim4pt{\cal Z}^{\rm short}_{\yng(3), \rm suprim} &=&
\frac{y\, (\frac{1}{a^2 b^2}-\frac{c}{a}-\frac{d}{b})}{(1-a^2
b^2)\,(1-a^3b^3)}\;,
\nn\\[2ex]
\Yboxdim4pt{\cal Z}^{\rm short}_{\yng(1,1,1), \rm suprim} &=&
\frac{y\,a^3b^3\, (\frac{1}{a^2
b^2}-\frac{c}{a}-\frac{d}{b})}{(1-a^2 b^2)\,(1-a^3b^3)}\;,
\nn\\[2ex]
\Yboxdim4pt{\cal Z}^{\rm short}_{\yng(4), \rm suprim} &=&
\frac{y^2\,(1+cd\,(a^3b^3+a^5b^5+a^8b^8)+c^2a^7b^9+d^2a^9b^7)}
{a^2b^2\,(1-a^2b^2)\,(1-a^3b^3)\,(1-a^4b^4)}
\nn\\[1ex]
&&{} -\frac{y^2\,(cb+da)(a^2b^2+a^3b^3+a^4b^4-a^6b^6)}
{(1-a^2b^2)\,(1-a^3b^3)\,(1-a^4b^4)} \;,
\nn\\[2ex]
\Yboxdim4pt{\cal Z}^{\rm short}_{\yng(2,2), \rm suprim} &=&
\frac{y^2\,(1+cd\,(\frac1{ab}+a^2b^2+a^3b^3)+c^2ab^3+d^2a^3b-(cb+da)\,(1+ab))}
{(1-a^2b^2)^2(1-a^3b^3)} \;,
\nn\\[2ex]
\Yboxdim4pt{\cal Z}^{\rm short}_{\yng(2,1,1), \rm suprim} &=&
\frac{y^2(ab+cd\,(1+ab+a^2b^2)+c^2b^2+d^2a^2-(cb+da)\,(\frac1{ab}+a^2b^2))}
{(1-ab)(1-a^2b^2)(1-a^4b^4)} \;.
%\nn\\[2ex]
%{\cal Z}^{\rm short}_{4, \rm suprim}
%&=&y^2 {1+2 a^3 b^3+a^4 b^4\over a^2 b^2 (1-a^2 b^2)^2(1-a^4 b^4)}\nn\\
%&&{}
% +\frac{y^2\,(b^2 c^2\!+\! a^2 d^2)\,(1\!+\!2ab\!+\!a^4b^4)}
% {(1-a^2b^2)^2(1-a^4b^4)}
% +\frac{y^2 cd\, (1+ 3 a^2b^2 + a^3b^3 + a^5b^5) }
% {ab\,(1-ab)(1-a^2b^2)(1-a^4b^4)}
% \nonumber\\
% &&{}
% -\frac{y^2\,(cb+da)}{ab\,(1-ab)^2}
% -\frac{2y^2\,(cb+da)\,a^3b^3}{ab\,(1-ab)^2(1-a^4b^4)}
% \;,\nn\\
% \mbox{etc.}
\nn
\end{eqnarray}
Continuing to higher $L$, the complete list of semishort
multiplets appearing in the ${\cal N}=4$ SYM spectrum is obtained.
The $\alSU(2)^2\times \alSU(4)$ charges can be read off from
(\ref{weight}) i.e. $[{n_a\over 2},{n_b\over
2}][n_c,n_y-n_c-n_d,n_d]$, while
 \[
 \Delta=2+n_y+\ft12(n_a+n_b) ,\quad\quad\quad~~~~  B=\ft12(n_d-n_c)
 \;,
 \]
 and $L$ is specified by the subscript of ${\cal Z}$'s.
The results for $L=2,3 $ precisely match~\eqref{123} with the
coefficients $c_n$ given by the expansion
\begin{eqnarray}
\sum_{n=0}^\infty c_n\, x^n=\frac{1}{(1\!-\!x^2)(1\!-\!x^3)}\;.
\label{cnco}
\end{eqnarray}
Notice that representatives for a given YT-pleton can be always
chosen inside the $\alhs(1,1)$ subgroup. This corresponds to
setting $c=d=0$ in (\ref{Zresult}).

 Similar techniques can be applied to the study of any
closed subsector in the SYM spectrum. For example $SU(4)$ singlets
in the $AC$ series are described by states saturating one of the
bounds (\ref{BPS}) and $q_1=p=q_2=0$. The various conditions for
the first bound combine to give
\[
n_{c_1}= n_{c_2}=n_{a_2}=0  \quad n_{d_1}=n_{d_2} =  L  \;.
\]
This leads to the $\alSU(2,1)$ invariant subsector with letters
${a_1^\dagger}^{2+n}\, {b_1^\dagger}^{m}\,
{b_2^\dagger}^{n-m}d^\dagger_1 d^\dagger_2|Z\rangle$, which are
essentially derivatives of the self-dual field strength
$D_{11}^{m} D_{12}^{n-m} F_{11}$. Anomalous dimensions for
three-letter states of this type will be computed in the next
section using the corresponding $\alSU(1,2)$ spin chain.

\subsubsection*{Semishort multiplets group into long multiplets}
We can now explicitly show that the semi-short multiplets
appearing in the free ${\cal N}=4$ SYM spectrum above organize
into long multiplets. This is expected since after switching on
interactions the shortening conditions~(\ref{eq:shortening}) are
generically no longer satisfied. Specifically, a semishort
multiplet appearing in the decomposition of an $L$-pleton joins
two multiplets from the $(L\!+\!1)$-pleton and a fourth one from
the $(L\!+\!2)$-pleton to build a long multiplet according
to~(\ref{ltoshort}). The semishort multiplets appearing in this
decomposition are related to each other by the action of $Q^1_{-}$
and $\bar{Q}_{4-}$ in (\ref{qs}).

Our statement then is equivalent to claiming that the total
partition function of the semi-short SYM spectrum
\begin{eqnarray} {\cal Z}^{\rm rst} &=& \sum_{n=2}^\infty
 {\cal Z}^{\rm rst}_n \;,
\end{eqnarray}
after subtraction of the $\ft12$-BPS multiplets contains the
factors $(1-y \frac{c}{a})$ and $(1-y\, \frac{d}{b})$. To prove
this, we write the total partition function as
\begin{eqnarray}
{\cal Z}^{\rm rst}&=&  - (1-a b)\sum_{k=1}^\infty
\frac{\varphi(k)}{k} \ln \Big\{ 1-y^k \left( \frac{1+
(cd)^k-(ad)^k-(bc)^k}{1-(ab)^k}\right)
 \Big\}\nn\\[.5ex]
&& - y \,(1+cd-ad-bc) \;,
\end{eqnarray}
while for the total partition function of $\ft12$-BPS multiplets
we obtain
\[ {\cal Z}^{\rm
rst}_{\frac12-{\rm BPS}} ~=~ \sum_{n=2}^\infty  \,{\cal V}^{\rm
rst,n}_{[0^\dagger,0^\dagger][0,n,0]} = \,
\frac{y^2(1-ad)(1-bc)}{(1-y)(1-cd)}- \frac{(cdy)^2\,
(a-c)(b-d)}{(1-cd)(1-y c d)} \;.
\]
Using
\[
\sum_{k=1}^\infty \frac{\varphi(k)}{k} \ln (1-x^k)=-\frac{x}{1-x}
\;,
\]
one finds indeed that
\[
\big({\cal Z}^{\rm rst}|_{y\to \frac{a}{c}}- {\cal Z}^{\rm
rst}_{\frac12-{\rm BPS}}\big)\Big|_{y\to \frac{a}{c}} ~=~0\;,
\]
and likewise for $y\to \frac{b}{d}$. Hence, the semishort
multiplets in the free SYM spectrum organize in long multiplets
whose highest weight states are collected in the regular function
\begin{eqnarray} {\cal Z}^{\rm long}_{\rm suprim} &\equiv&
\frac1{y^2\,abcd}\,T^{-1}_{\rm long} \,\left( {\cal Z}^{\rm
rst}-{\cal Z}^{\rm rst}_{\frac12{\rm BPS}} \right)_{\rm HW}
+\frac{1}{a^2 b^2}\,\frac1{1-y}\;,
\end{eqnarray}
with
\[
T_{\rm long}=(1-y \frac{c}{a}) (1-y \frac{d}{b})(1-ad)(1-bc)(1-
\frac{a}{c})(1-\frac{b}{d}) \;, \label{klm}
 \]
defining the restriction of the long Konishi multiplet.

%%%%%%%%%%%%%%%%%%%%%%%%%%%%%%%%%%%%%%%%%%%%%%%

\section{Symmetry breaking and anomalous dimensions}
\label{sanodim}

In the interacting theory only one out of the infinite tower of
conserved current doubleton multiplets
\[
\label{tksdjf} {\cal Z}_{\Yboxdim4pt \yng(2)}=\sum_{n=0}^\infty
\mult_{2n}, \qquad \mult_{j}:=\mult^{j,0
}_{[-1+\frac{1}{2}j^\ast,-1+\frac{1}{2}j^\ast][0,0,0]}.\]
is protected against quantum corrections to the scaling dimension:
the $\superN=4$ supercurrent multiplet
$\mult_0=\mult^{2,0}_{[0^\dagger,0^\dagger][0^\dagger,2,0^\dagger]}$.
The remaining multiplets $\mult_{2n}$ acquire anomalous dimensions
which violate the conservation of their HS currents at the quantum
level. At one-loop, one has
\cite{Kotikov:2000pm,Kotikov:2001sc,Dolan:2001tt}
\[
\gamma_{\rm 1-loop}(2n) =\frac{\gym^2 N}{2\pi^2}\,h(2n),\qquad
h(j)=\sum_{k=1}^j\frac{1}{k},
\]
This elegant ('number theoretic') formula gives a clue on how to
compute generic anomalous dimensions at first order in
perturbation theory relying on symmetry breaking considerations.
Naively, one would look for all occurrences of the broken currents
$\mult_{2n}$ within some operator $\mathcal{O}$. Each occurrence
of some broken current should contribute to the anomalous
dimension of $\mathcal{O}$ a term proportional to $h(2n)$. Indeed,
this is nearly what happens, the one-loop dilatation operator
\cite{Beisert:2003jj} can be written as
\[
H=\sum_{s=1}^L H_{(s,s+1)} = \sum_{s=1}^L \sum_{j=0}^\infty 2h(j)
\,P^{j}_{(s,s+1)} ,
\]
where $P^{j}_{(s,s+1)}$ projects the product of fields (`letters')
at nearest neighboring sites $s$ and $s+1$ onto $\mult_{j}$. Here,
the sum goes over all values of $j$ and not just the even ones.
The point is that although bilinear currents $\mult_{2n+1}$
corresponding to the broken generators are eliminated after
tracing over color indices, they still  appear in subdiagrams
\,${\Yboxdim5pt \yng(1,1)}$\, inside a bigger trace. The
corresponding decomposition for doubletons is given in
\eqref{123}.

%%%%%%%%%%%%%%%%%%%%%%%%%%%%%%%%%%%%%%%%%%%%%%%%%%%%%%%%%%%%%%%%%%%%%%%%%%%%%%

\subsection{`Twist' three anomalous dimensions}

Expressions \eqref{Zresult} give not only multiplicities and
charges of semishort primaries but also a representative of each
multiplet in terms of the oscillators $(a_1,b_1,c_2,d_1)$. For
instance states in the $\alSL(2)$ sector inside $\alPSU(1,1|2)$,
associated to words made out of powers of $a_1 b_1$ (i.e.~a single
scalar and all its derivatives) can be taken as representatives
for semishort multiplets $\mult_{0,n}$. The letters in this sector
are:
\[
|k\rangle_0=(a_1^\dagger b_1^\dagger )^k|Z\rangle \quad
\leftrightarrow \quad
 {\cal D}_1^{k} Z \quad ,
\]
with ${\cal D}_i=D_{\alpha=i,\dot{\alpha}=i}$. Similarly
derivatives of $ (a^\dagger d^\dagger)^3$ appearing in
$Q_1^+\bar{Q}_4^+ T_{\rm short}\,{\cal Z}^{\rm rst}_{3, \rm
suprim} $ can be chosen as representatives of $\mult_{1,n}$
(similarly for the conjugate multiplets $\mult_{-1,n}$). Indeed
there is a single state of this type inside each fermionic
semishort multiplet in ${\cal Z}_3^{\rm rst}$. The letters are
now:
\[
|k\rangle_1=(a_1^\dagger b_1^\dagger)^k (a_1^\dagger
d_1^\dagger)|Z\rangle \quad \leftrightarrow \quad {\cal
D}_1^{k}\lambda \quad .
\]
with $\lambda=\lambda_{\alpha=1,\dot{r}=1}$ one of the gaugini.
 The $\alSL(2)$ generators in both cases can be written as
\[
J_-=a_1b_1\;, \quad\quad J_+=a_1^\dagger b_1^\dagger\;, \quad
\quad J_3=\ft12 a_1 a_1^\dagger- \ft12 b_1^\dagger b_1\;,
\]
while the spin is given by
$J_3|k\rangle_m=(\ft12+\frac{m}{2})|k\rangle_m$. Therefore for
$\mult_{m,n}$, $m=0,1$, we use the $\alSL(2)$ spin chain with spin
$\half+\half m$. We use a unified notation for a single spin state
of either chains
\[\label{safdg}
\state{k}_m=(a_1^\dagger)^{k+m}(b_1^\dagger)^{k}
(d_1^\dagger)^{m}|Z\rangle \quad \leftrightarrow \quad m{\cal
D}_{1}^{k} \lambda + (1-m) {\cal D}_{1}^{k} Z \quad .
\]
The Hamiltonian of the relevant (super) spin chains in the two
subsectors can be computed using the harmonic action in
\cite{Beisert:2003jj}. The resulting Hamiltonian `density' is
\[\label{kjdshg}
H_{(12)} \state{k,n-k}_m=\sum_{k'=0}^n c^{(m)}_{n,k,k'}
\state{k',n-k'}_m \;,
\]
with coefficients
\[\label{oirjxdsulm}
c^{(m)}_{n,k,k'}=
\begin{cases}
h(k+m)+h(n-k+m)&\mbox{for }k=k'\\
\displaystyle\frac{k!(n-k+m)!}{k'!(n-k'+m)!(k-k')}&\mbox{for }k>k',\\
\displaystyle\frac{(n-k)!(k+m)!}{(n-k')!(k'+m)!(k'-k)}&\mbox{for
}k<k'.
\end{cases}
\]
For $m=0$ this is equivalent to the $\alSL(2)$ subsector of
letters $\cder^n Z$ up to a rescaling by $n!$

For multiplets $\mult_{m,n}$ with $m\geq 2$ we use the
$\alSU(1,2)$ spin chain corresponding to the closed subsector with
residual symmetry algebra $\alSU(1,2)$. The spin states are now
specified by two conserved charges corresponding to the rank two
algebra $\alSU(1,2)$
\[\label{sdlfkjafdsa}
\state{k,l}=(a_1^\dagger)^{2+k+l}(b_{1}^\dagger)^{k}(b_{2}^\dagger)^{l}
d_1^\dagger d_2^\dagger|Z\rangle  \quad \leftrightarrow \quad
{\cal D}_{1}^{k} {\cal D}_{2}^{l} {\cal F} \quad ,
\]%
with ${\cal F}=F_{\alpha=1,\beta=1}$. The planar, one-loop
dilatation generator $H$ acts on two adjacent spin sites as
\[\label{sdlfkjafdsadd}
H_{(12)} \state{k,l;m-k,n-l}= \sum_{r=0}^m\sum_{s=0}^n
c^{m,n}_{k,l;r,s} \state{r,s;m-r,n-s} \;,
\]%
with
\[ \label{lwkrgre}
c^{m,n}_{k,l;r,s}=\begin{cases} h(2+k+l)+h(2+n+m-k-l),
  &\mbox{for } k=r,l=s,\\
\displaystyle
-\frac{k!l!(2+n+m-k-l)!(k+l-r-s-1)!}{r!s!(2+n+m-r-s)!(k-r)!(l-s)!},
  &\mbox{for }k\geq r,l\geq s,\\
\displaystyle
-\frac{(m-k)!(n-l)!(2+k+l)!(r+s-k-l-1)!}{(m-r)!(n-s)!(2+r+s)!(r-k)!(s-l)!},
  &\mbox{for }k\leq r,l\leq s,\\
0,&\mbox{for }k>r,l<s,\\
0,&\mbox{for }k<r,l>s.
\end{cases}
\]
The coefficients again follow from the harmonic action.

\begin{table}\centering
$\begin{array}[t]{|c|ccccccc|}\hline n\backslash m &0&1&2&3&4&5&6
\\\hline
3&\frac{15}{16} & \frac{5}{4} & \frac{47}{32} & \frac{131}{80} &
\frac{71}{40} & \frac{1059}{560} & \frac{4461}{2240} \\
5&\frac{35}{32} & \frac{133}{96} & \frac{761}{480} &
\frac{487}{280} &
\frac{12533}{6720} & \frac{39749}{20160} & \frac{13873}{6720} \\
6&\frac{227}{160} & \frac{761}{480} & \frac{967}{560} &
\frac{2069}{1120}
& \frac{39349}{20160} & \frac{2747}{1344} & \frac{3929}{1848} \\
7&\frac{581}{480} & \frac{179}{120} & \frac{3763}{2240} &
\frac{18383}{10080} & \frac{39133}{20160} & \frac{7543}{3696} &
\frac{94373}{44352} \\
8&\frac{5087}{3360} & \frac{1403}{840} & \frac{18187}{10080} &
\frac{38677}{20160} & \frac{49711}{24640} & \frac{2593}{1232} &
\frac{629227}{288288} \\
\hline
\end{array}$
\caption{First few paired anomalous dimensions for $\mult_{m,n}$
with $L=3$} \label{tab:Twist3b}
\end{table}
We now compute the spectrum of one-loop planar anomalous
dimensions explicitly using
(\ref{kjdshg})--(\ref{lwkrgre}).\footnote{The Hamiltonian is
related to the dilatation operator by $\delta
D=(\gym^2N/8\pi^2)H+\order{\gym^3}$.} By inspecting the spectrum
of lowest-lying states and their energies we find that almost all
of them form pairs with degenerate energies.
We list the pairs in \tabref{tab:Twist3b}.%
\footnote{The energies are all rational numbers because there is
always just a single pair up to $n\leq 8$. Starting from $n=9$
there is more than one pair and the energies become irrational.}
\begin{table}\centering
$\begin{array}[t]{|c|ccccccc|}\hline n\backslash m &0&1&2&3&4&5&6
\\\hline
0&            0 & \frac{3}{4} & \frac{9}{8} & \frac{11}{8} &
\frac{25}{16}
& \frac{137}{80} & \frac{147}{80} \\
2&  \frac{1}{2} & \frac{25}{24} & \frac{4}{3} & \frac{123}{80} &
\frac{407}{240} & \frac{3067}{1680} & \frac{271}{140} \\
4&  \frac{3}{4} & \frac{49}{40} & \frac{71}{48} & \frac{929}{560}
&
\frac{9}{5} & \frac{9661}{5040} & \frac{2259}{1120} \\
6&\frac{11}{12} & \frac{761}{560} & \frac{191}{120} &
\frac{8851}{5040} &
\frac{66}{35} & \frac{221047}{110880} & \frac{21031}{10080} \\
8&\frac{25}{24} & \frac{7381}{5040} & \frac{101}{60} &
\frac{101861}{55440} & \frac{6581}{3360} & \frac{329899}{160160} &
\frac{21643}{10080} \\
10&\frac{137}{120} & \frac{86021}{55440} & \frac{493}{280} &
\frac{2748871}{1441440} & \frac{20383}{10080} &
\frac{203545}{96096} &
\frac{122029}{55440} \\
\hline
\end{array}$
\caption{First few unpaired anomalous dimensions for $\mult_{m,n}$
with $L=3$. Parity is given by $(-1)^{m+1}$.} \label{tab:Twist3a}
\end{table}
For the unpaired states one can observe a pattern in the table of
energies, \tabref{tab:Twist3a}. We find that all energies agree
with the formula
\[
\delta D=\frac{\gym^2 N}{8\pi^2} \bigbrk{ 2h(\half m-\half) +2
h(m+\half n) +2 h(\half m+\half n) -2 h(-\half) } \;.
\]
In particular, for $m=1$ the energies are
\[
\delta D=\frac{\gym^2 N}{8\pi^2} \bigbrk{ +2 h(1+\half n) +2
h(\half+\half n) -2 h(-\half)} =\frac{\gym^2 N}{2\pi^2}\,h(n+2)
\;,
\]
which agrees precisely with the energy of the short twist 2
multiplet $\mult_{2n+2}$, \eqref{tksdjf}. Superconformal
invariance requires this degeneracy so that the short multiplets
can join to form a long multiplet. The cases $m=0$ and $n=0$ also
seem interesting, we find $\delta D=(\gym^2 N/8\pi^2) 4h(\half n)$
and $\delta D=(\gym^2 N/8\pi^2) 6 h(m)$.

Let us note a peculiarity of the three parton states discussed
above. Intriguingly, for $\mult_{m,n}$ we can reproduce all
$\alSU(2,1)$ spin chain results also with a $\alSL(2)$ spin chain
with spin $-m/2-1/2$ and $n$ excitations given by
(\ref{kjdshg}),(\ref{oirjxdsulm}).

\section{Conclusions}
\label{sconclusions}

In the present paper, we have studied the decomposition of the
spectrum of single-trace gauge invariant operators of free ${\cal
N}=4$ SYM theory with $SU(N)$ gauge group in irreps
of~$\alhs(2,2|4)$, the HS extension of the superconformal
algebra~$\alPSU(2,2|4)$. To this end we have shown that HS
$L$-pleton multiplets can be associated to Young tableaux made of
$L$ boxes, each representing a singleton of $\alPSU(2,2|4)/
\alhs(2,2|4)$, compatible with the cyclicity of the trace over
color indices. For other gauge groups, further restrictions are to
be imposed. For $L=2$ only the symmetric product gives rise to
physical operators independently of the choice of the (simple)
gauge group~\cite{Sezgin:2001zs}. The antisymmetric doubleton is
ruled out by the cyclicity of the trace but still its
decomposition is relevant to diagrammatic computations of
composite operators where such combinations appear in intermediate
channels. We have then focussed on tripletons associated to Young
tableaux with $L=3$ boxes. The only tableaux compatible with the
cyclicity of the trace are the totally symmetric ($d_{abc}$) and
antisymmetric ($f_{abc}$) tripletons. The former includes the KK
recurrences of the doubleton and the latter part of the Goldstone
fields. The remaining Goldstone fields belong (in the free theory)
to the $L=4$-letter `window' $\Yboxdim6pt \yng(2,2)$.

For higher $L$-pletons we have identified all operators belonging
to BPS or semishort-semishort multiplets of $\psu$ in the free
theory. In particular, we have derived the partition function for
${\cal N}=4$ superconformal primaries saturating both left and
right unitarity bounds. After interactions are turned on, they are
shown to combine such as to give rise to long multiplets of the
superconformal group as expected from the boundary description of
the `Grande Bouffe' in the AdS bulk.

Finally, we have computed anomalous dimensions of operators that
appear in the decomposition of tripletons in terms of $\psu$
multiples. Remarkably the resulting anomalous dimensions for the
full tripleton tower follow from integrable spin chains with
symmetry group $\alSL(2)_j$ and arbitrarily high spin $j$. The
regularity of the pattern suggests the presence of some not-so
`hidden' symmetry. Indeed there are by now various independent
indications that some aspects of the dynamics of large $N$ ${\cal
N}=4$ SYM theory and its holographic dual type IIB superstring on
$AdS_5 \times S^5$ expose an integrable structure. In the latter,
the supercoset structure of the target superspace and the
(generalized) flatness of the supercoset currents allow one to
identify an infinite number of conserved charges that form a
Yangian \cite{Bena:2003wd, Alday:2003zb, Vallilo:2003nx}. In the
former, the dilatation operator can be identified with the
Hamiltonian of an integrable super-spin chain to lowest orders.
Some of the infinite charges have been identified and given
explicit perturbative expressions. These two routes to
integrability have been connected in \cite{Dolan:2003uh}. In the
emergence of the integrable structure HS symmetry enhancement has
so far played only a marginal role. Yet HS dynamics in lower
dimensions is typically formulated in terms of a Cartan integrable
system
\cite{Vasiliev:2003ev,Alkalaev:2002rq,Vasiliev:2001wa,Sezgin:2001zs,Sezgin:2001yf,Sezgin:2002rt}.
It is then tempting to speculate that at least at one loop and
large $N$, HS symmetry could explain the pattern of mass-shifts
and anomalous dimensions and give some additional insight into the
geometric origin of integrability.

\section*{Acknowledgements}

This work was supported in part by I.N.F.N., by the EC programs
HPRN-CT-2000-00122, HPRN-CT-2000-00131 and HPRN-CT-2000-00148, by
the INTAS contract 99-1-590, by the MURST-COFIN contract
2001-025492 and by the NATO contract PST.CLG.978785. N.B. dankt
der \emph{Studienstiftung des deutschen Volkes} f\"ur die
Unterst\"utzung durch ein Promotions\-f\"orderungsstipendium.

\appendix
\section{${\cal N}=4$ shortening, yet again.}
\label{sa1}

Here we collect some notation for representations of the ${\cal
N}=4$ superconformal
 algebra $\psu$ and their shortenings.
We denote by
\<\label{V}\mult^{\Delta,B}_{[j,\jb][q_1,p,q_2]}\;, \label{VA}\>
a generic long multiplet of $\alPSU(2,2|4)$ with highest weight
state in the ${\cal R}_{[j,\jb][q_1,p,q_2]}$ representation of
$\alSU(2)\times \alSU(2)\times \alSU(4)$, conformal dimension
$\Delta$ and hypercharge $B$. As above, $[q_1,p,q_2]$ are Dynkin
labels of $\alSU(4)$ while $[j,\jb]$ denote the spins under
$\alSU(2)\times \alSU(2)$.

The representation content of the long multiplet \eqref{VA} under
the bosonic subalgebra $\mathfrak{su}(2)\times
\mathfrak{su}(2)\times\mathfrak{su}(4)$ may be found from
evaluating the tensor product~$\mult^{2,0,0}_{[0,0][0,0,0]}
\chi_{[j,\jb][q_1,p,q_2]}^{\Delta-2,B,P}$, with the long Konishi
multiplet $\mult^{2,0,0}_{[0,0][0,0,0]}$, or explicitly by using
the Racah-Speiser algorithm as
\begin{eqnarray} \mult^{\Delta,B}_{[j,\jb][q_1,p,q_2]}
 &=&
\sum_{\epsilon_{A\alpha},\bar{\epsilon}^A_{\dot{\alpha}}\in
\{0,1\}} {\chi}_{[j,\jb][q_1,p,q_2]+ \epsilon_{A\alpha} {\cal
Q}^A_{\alpha}
 + \bar{\epsilon}^A_{\dot{\alpha}}
 \bar{{\cal Q}}_{A\dot{\alpha}}}\;,
 \label{susy}
 \end{eqnarray}
with the sum running over the $2^{16}$ combinations of the $16$
supersymmetry charges ${\cal Q}^A{}_{\alpha}$, $\bar{\cal
Q}_{A\dot{\alpha}}$, $A=1, \dots, 4,\alpha,\dot{\alpha}=1,2$ with

Dynkin labels\footnote{Notice the flip of notations for the
conjugate charges with respect to~\cite{Bianchi:2003wx} and the
unconventional use of oscillators in the denominator to mean
annihilation operators.}
\begin{eqnarray}
&&{\cal Q}^1{}_{\alpha} = a^\dagger_{\alpha}\,{c_1}\equiv
\frac{a_{\alpha}}{c_1}=[\pm \ft12,0][1,0,0] \;, \quad
\bar{\cal Q}_{1\dot{\alpha}}=b^\dagger_{\dot{\alpha}}\,c^\dagger_1 \equiv b_{\dot{\alpha}}\,c_1=[0,\pm\ft12][-1,0,0]\nn\\
&&{\cal Q}^2{}_{\alpha} = a^\dagger_{\alpha}\,{c_2}\equiv
\frac{a_{\alpha}}{c_2}=[\pm \ft12,0][-1,1,0] \;,\quad \bar{\cal
Q}_{2\dot{\alpha}}=b^\dagger_{\dot{\alpha}}\,c^\dagger_2\equiv
b_{\dot{\alpha}}\,c_2=[0,\pm\ft12][1,-1,0]
\;,\nn\\
&&{\cal Q}^3{}_{\alpha} = a^\dagger_{\alpha} \,d^\dagger_1\equiv
a_{\alpha} \,d_1=[\pm \ft12,0][0,-1,1] \;,\quad \bar{\cal
Q}_{3\dot{\alpha}}= b^\dagger_{\dot{\alpha}}\,d_1\equiv
\frac{b_{\dot{\alpha}}}{d_1}=[0,\pm\ft12][0,1,-1]
\;,\nn\\
&&{\cal Q}^4{}_{\alpha} = a^\dagger_{\alpha} \,d^\dagger_2\equiv
a_{\alpha} \,d_2=[\pm\ft12,0][0,0,-1] \;,\quad \bar{\cal
Q}_{4\dot{\alpha}}=b^\dagger_{\dot{\alpha}}\,d_2\equiv
\frac{b_{\dot{\alpha}}}{d_2}=[0,\pm\ft12][0,0,1] \;. \label{qs}
\end{eqnarray}
Every ${\cal Q}^A{}_{\alpha}$, $\bar{\cal Q}_{A\dot{\alpha}}$
 raises the conformal dimension by $\ft12$,
parity is left invariant, and the hypercharge $B$ is lowered and
raised by $\ft12$ by each ${\cal Q}^A{}_{\alpha}$ and $\bar{\cal
Q}_{A\dot{\alpha}}$ respectively. In order to make sense out of
\eqref{susy} also for small values of $q_1, p, q_2, j, \jb$, we
note that the character polynomials associated with negative
Dynkin labels are defined according to
\begin{eqnarray} {\chi}_{[j,\jb][q_1,p,q_2]}&=&
-{\chi}_{[j,\jb][-q_1-2,p+ q_1+ 1,q_2]} ~=~
-{\chi}_{[j,\jb][q_1,p+ q_2+1,- q_2-2]} \non &=&
-{\chi}_{[j,\jb][q_1+p+1,-p-2,q_2+p+1]} \non &=&
-{\chi}_{[-j-1,\jb][q_1,p,q_2]} ~=~
-{\chi}_{[j,-\jb-1][q_1,p,q_2]}
 \;.
\la{negw} \end{eqnarray}
In particular, this implies that ${\chi}_{[j,\jb][q_1,p,q_2]}$ is
identically zero whenever any of the weights $q_1$, $p$, $q_2$
takes the value~$-1$ or one of the spins $j$, $\jb$ equals
$-\ft12$.

In $\superN=4$ SYM, there are two types of (chiral) shortening
conditions for particular values of the conformal dimension
$\Delta$: BPS (B) which may occur when at least one of the spins
is zero, and semi-short (C) ones. The corresponding multiplets are
constructed similar to the long ones \eqref{susy}, with the sum
running only over a restricted number of supersymmetries.
Specifically, the critical values of the conformal dimensions and
the restrictions on the sums in \eqref{susy} are given by
\<\label{eq:shortening}
\begin{tabular}{llll}
B$_L$: \quad&
$\mult^{\Delta,B}_{[0^\dagger,\jb][q_1,p,q_2]}$\qquad\qquad &
$\Delta= p+\sfrac{3}{2}q_1+\sfrac{1}{2}q_2$ & $\epsilon_{1\pm}=0$
\\[1ex]
B$_R$: & $\mult^{\Delta,B}_{[j,0^\dagger][q_1,p,q_2]}$& $\Delta=
p+\sfrac{1}{2}q_1+\sfrac{3}{2}q_2$ & $\bar{\epsilon}_{4\pm}=0$
\\[1ex]
C$_L$: & $\mult^{\Delta,B}_{[j^*,\jb][q_1,p,q_2]}$&
$\Delta=2+2j+p+\sfrac{3}{2}q_1+\sfrac{1}{2}q_2$ &
$\epsilon_{1-}=0$
\\[1ex]
C$_R$: & $\mult^{\Delta,B}_{[j,\jb^*][q_1,p,q_2]}$ &
$\Delta=2+2\jb+p+\sfrac{1}{2}q_1+\sfrac{3}{2}q_2$ \qquad\qquad&
$\bar{\epsilon}_{4-}=0$
 \end{tabular}
 \>
for the different types of multiplets. They represent the basic
$\ft18$-BPS and $\ft{1}{16}$ semishortenings in ${\cal N}=4$ SCA
and are indicated as in with a ``$\dagger$'' and a ``$*$''
respectively.

If the conformal dimension $\Delta$ of the HWS of a long multiplet
\eqref{VA} satisfies one of the conditions~\eqref{eq:shortening},
the multiplet splits according to
\<\label{eq:splitting0} \mathrm{L}:\quad
\mult^{\Delta,B}_{[j,\jb][q_1,p,q_2]}\eq
\mult^{\Delta,B}_{[j^\ast,\jb][q_1,p,q_2]}+ \mult^{\Delta+
\frac{1}{2},B-\frac{1}{2}}_{[j-
\frac{1}{2}^\ast,\jb][q_1+1,p,q_2]}\;, \quad \nln \mathrm{R}:\quad
\mult^{\Delta,B}_{[j,\jb][q_1,p,q_2]}\eq
\mult^{\Delta,B}_{[j,\jb^\ast][q_1,p,q_2]}+
\mult^{\Delta+\frac{1}{2},B+ \frac{1}{2}}_{[j,\jb-
\frac{1}{2}^\ast][q_1,p,q_2+1]}\;, \>
where by `$^*$' we denote the $1/16$ semishortening. Consequently,
we denote by $\mult^{\Delta,B}_{[j^\ast,\jb^\ast][q_1,p,q_2]}$ the
$1/8$ semi-short multiplets appearing in the decomposition
\begin{eqnarray} \mult^{\Delta,B}_{[j,\jb][q_1,p,q_2]}\eq
\mult^{\Delta,B}_{[j^\ast,\jb^\ast][q_1,p,q_2]}+ \mult^{\Delta+
\frac{1}{2},B-\frac{1}{2}}_{[j-
\frac{1}{2}^\ast,\jb^\ast][q_1+1,p,q_2]}+
\mult^{\Delta+\frac{1}{2},B+ \frac{1}{2}}_{[j^\ast,\jb-
\frac{1}{2}^\ast][q_1,p,q_2+1]}\non &&{} +
\mult^{\Delta+1,B}_{[j-\frac12^\ast,\jb-
\frac{1}{2}^\ast][q_1+1,p,q_2+1]} \;,
\label{ltoshort}\end{eqnarray}
if left and right shortening conditions in \eqref{eq:shortening}
are simultaneously satisfied. The semishort multiplets appearing
in this decomposition are constructed explicitly according to
\eqref{susy}, \eqref{eq:shortening}.

Formulae \eqref{eq:splitting0} include the special cases
$\mult^{\Delta,B}_{[j^\ast,\jb][0,p,q_2]}$,
$\mult^{\Delta,B}_{[j^\ast,\jb][0,0,q_2]}$, and
$\mult^{\Delta,B}_{[j^\ast,\jb][0,0,0]}$, corresponding to
(chiral) $1/8$, $3/16$, and $1/4$ semi-shortening, respectively;
likewise for $\mult^{\Delta,B}_{[j,\jb^\ast][q_1,p,0]}$,
$\mult^{\Delta,B}_{[j,\jb^\ast][q_1,0,0]}$, and
$\mult^{\Delta,B}_{[j,\jb^\ast][0,0,0]}$. For $j=0$ and $\jb=0$,
respectively, the decompositions~\eqref{eq:splitting0} yield
negative spin labels. They are to be interpreted as BPS
multiplets, denoted by `$^\dagger$', as follows
\begin{eqnarray}
\mult^{\Delta,B}_{[- \frac{1}{2}^\ast,\jb][q_1,p,q_2]} \equiv
\mult^{\Delta+ \frac{1}{ 2},B+
\frac{1}{2}}_{[0^\dagger,\jb][q_1+1,p,q_2]}\;, \qquad
\mult^{\Delta,B}_{[j,- \frac{1}{2}^\ast][q_1,p,q_2]} \equiv
\mult^{\Delta+ \frac{1}{2},B-
\frac{1}{2}}_{[j,0^\dagger][q_1,p,q_2+1]}\label{bps} \;,
\end{eqnarray}
where one verifies that the BPS highest weight states satisfy the
BPS shortening conditions of \eqref{eq:shortening}. In addition,
there is the series
$\mult^{p,0}_{[0^{\dagger},0^{\dagger}][0^\dagger,p,0^\dagger]}$
of $\ft12$-BPS multiplets.

For convenience (but not quite accurately) we can also define
\[
\mult^{p,0}_{[-1^\ast,-1^\ast][0,p,0]} :=
\mult^{p+2,0}_{[0^{\dagger},0^{\dagger}][0^\dagger,p+2,0^\dagger]}.
\label{bpsnot}
\]

\section{Oscillator description}
\label{sao} Here we collect some useful formulae, concerning the
oscillator description of $\psu$ representations.

\subsection{$\alSU(2)\times\alSU(2|4)$ invariant vacuum}

The unphysical $\alSU(2)\times\alSU(2|4)$ invariant vacuum
$|U\rangle$ is defined as the ground state of the set of bosonic
${a}_{\alpha}^{(s)},{b}^{(s)}_{\dot \alpha}$ and fermionic
oscillators ${\theta}_{A}^{(s)}$:
\[
{a}_{\alpha,i}|U\rangle ={b}_{\dot
\alpha}^{(s)}|U\rangle={\theta}_{A}^{(s)}|U\rangle=0 \;,
\]
with the vector index $s=1,\dots, L$ running over the sites in the
SYM state and ${\alpha},{\dot \alpha}=1,2$ $A=1,2,3,4$.
Oscillators satisfy the usual creation-annihilation commutation
relations:
\begin{eqnarray}  [ a_{\alpha}^{(s)},
a^{\beta}_{(s')} ]&=&\delta_{ss'} \delta_{\alpha}^\beta \quad\quad
[ b_{\dalpha }^{(s)} ,
b^{\dbeta}_{(s)}]=\delta_{ss'} \delta_{\dalpha}^{\dbeta}\;, \nn\\
 \{ \theta_{A }^{(s)}, \theta^{B}_{(s')} \} &=& \delta_{ss'}
\delta_{A}^B \;.   \end{eqnarray}

 A SYM state with $\psu$ charges $[q_1,p,q_2][j,\jb]^{\Delta,B,L}$
 can be constructed by
 acting on $|U\rangle$ with
 \begin{eqnarray}
\left(\begin{matrix} n_{a_1}\cr n_{a_2} \end{matrix} \right) &=&
\left(\begin{matrix}
 \ft12 \Delta+\ft12  B-\ft12 L+j\cr  \ft12 \Delta+ \ft12 B- \ft12 L-j
\end{matrix}\right)\;,\quad
\left(\begin{matrix} n_{b_1} \cr n_{b_2} \end{matrix} \right)=
\left(\begin{matrix} \ft12 \Delta-\ft12 B-\ft12 L+\jb\cr  \ft12
\Delta-\ft12 B-\ft12 L-\jb \end{matrix} \right)\;,\label{ns00wc}\nn\\
\left(\begin{matrix} n_{\theta_1} \cr n_{\theta_2} \cr
n_{\theta_3} \cr n_{\theta_4}
\end{matrix} \right) &=&
\left(\begin{matrix} \ft12 L-\ft12 B-\ft12 p-\ft34 q_1-\ft14
q_2\cr \ft12 L-\ft12 B-\ft12 p+\ft14 q_1-\ft14 q_2  \cr
 \ft12 L-\ft12 B+\ft12 p+\ft14 q_1-\ft14 q_2\cr
\ft12 L-\ft12 B+\ft12 p+\ft14 q_1+\ft34 q_2  \end{matrix} \right)
\;.
\end{eqnarray}
 The $\psu$ charges can instead be read from the inverse relations:
  \begin{eqnarray} \Delta&=& L+
\ft12 n_a+\ft12 n_b\;,\quad\quad
B = L- \ft12 n_{\theta} = \ft12 n_{a}-\ft12 n_{b}\;,\nn\\
\left[q_1,p,q_2\right] &=& \left[n_{\theta_2}-n_{\theta_1},
n_{\theta_3}-n_{\theta_2}
,n_{\theta_4}-n_{\theta_3} \right] \;,\nn\\[.5ex]
\left[j,\jb\right] &=&  \left[\ft12 (n_{a_1}- n_{a_2}),   \ft12
(n_{b_1}- n_{b_2}) \right] \;.\label{Ds0} \end{eqnarray}
Physical states are defined by the vanishing central charge
conditions:
\[
 {n}^{(s)}_a-{n}^{(s)}_b+{n}^{(s)}_\theta=2
 \;.
\label{zcond0}\] at every site $s=1,\ldots,L$.

\subsection{Physical vacuum}

The physical vacuum $|Z\rangle$ is defined as the ground state of
the set of $L$ species of bosonic
${a}^{(s)}_{\alpha},{b}^{(s)}_{\dot \alpha}$ and fermionic
oscillators ${c}_{r}^{(s)}=\theta_{r}^{(s)}$ and
${d}^{(s)}_{\dot{p}}= \bar\theta^{(s)}_{\dot{p}}$:
\[
{a}^{(s)}_{\alpha}|Z\rangle ={b}_{\dot\alpha}^{(s)}|Z
\rangle={c}_{p}^{(s)}|Z\rangle= { d}_{\dot{p}}^{(s)}|Z \rangle=0
\;,
\]
with the vector index $s=1,\dots, L$ running over the sites in the
SYM state and ${\alpha},{\dot\alpha},p,{\dot{p}}=1,2$. Oscillators
satisfy the usual creation-annihilation commutation relations:
\begin{eqnarray}
{} [ a_{\alpha}^{(s)}, a^{\beta}_{(s')} ]&=&\delta_{ss'}
\delta_{\alpha}^\beta\;, \quad\quad [ b_{\dalpha }^{(s)} ,
b^{\dbeta}_{(s')} ]=\delta_{ss'} \delta_{\dalpha}^{\dbeta} \;,\nn\\
 \{ c_{p}^{(s)}, c^{r}_{(s)} \} &=& \delta_{ss'}
\delta_{p}^r\;, \quad\quad \{ d_{\dot{p}}^{(s)}, d^{\dot{r}}_{(s)}
\}=\delta_{ss'} \delta_{\dot{p}}^{\dot{r}}\;.
\end{eqnarray}
A SYM state with $\psu$ charges $[q_1,p,q_2][j,\jb]^{\Delta,B,L}$
can be constructed by acting on $|ZZ\dots Z\rangle$ with
\begin{eqnarray}
\left(\begin{matrix} n_{a_1}\cr n_{a_2} \end{matrix} \right) &=&
\left(\begin{matrix}
 \ft12 \Delta+\ft12  B-\ft12 L+j\cr  \ft12 \Delta+ \ft12 B- \ft12 L-j
\end{matrix}\right)\;,\quad
\left(\begin{matrix} n_{b_1} \cr n_{b_2} \end{matrix} \right)=
\left(\begin{matrix} \ft12 \Delta-\ft12 B-\ft12 L+\jb\cr  \ft12
\Delta-\ft12 B-\ft12 L-\jb \end{matrix} \right)\;,\label{nsz}\\
\left(\begin{matrix} n_{c_1} \cr n_{c_2} \end{matrix} \right) &=&
\left(\begin{matrix} \ft12 L-\ft12 B-\ft12 p-\ft34 q_1-\ft14
q_2\cr \ft12 L-\ft12 B-\ft12 p+\ft14 q_1-\ft14 q_2  \end{matrix}
\right)\;, \quad \left(\begin{matrix} n_{d_1}\cr n_{d_2}
\end{matrix} \right) = \left(\begin{matrix} \ft12 L+\ft12 B-\ft12
p-\ft14 q_1+\ft14 q_2\cr \ft12 L+\ft12 B-\ft12 p-\ft14 q_1-\ft34
q_2
\end{matrix} \right)\;.
\nn
\end{eqnarray}
$\psu$ charges can instead be read from the inverse relations:
  \begin{eqnarray} \Delta&=& L+
\ft12 n_a+\ft12 n_b\;,\quad\quad
B = \ft12 n_{a}-\ft12 n_{b}\;,\nn\\
\left[q_1,p,q_2\right] &=& \left[n_{c_2}-n_{c_1},
L-n_{c_2}-n_{d_1}
,n_{d_1}-n_{d_2} \right] \;,\nn\\[.5ex]
\left[j,\jb\right] &=&  \left[\ft12 (n_{a_1}- n_{a_2}),   \ft12
(n_{b_1}- n_{b_2}) \right] \;. \label{Dsz} \end{eqnarray}
Physical states are defined by the vanishing central charge
conditions:
 \[
  C^{(s)} = {n}^{(s)}_a-{n}^{(s)}_b+{n}_c^{(s)}-{n}^{(s)}_d=0
\label{zcondz}
 \]
at every site $s=1,\ldots,L$.
\section{HS multiplets decomposition}
\label{a123}

For convenience of the reader we display here the translations of
formulae (\ref{123}). \< \Yboxdim4pt {\cal Z}_{\yng(2)}\eq
\sum_{n=0}^\infty\mult^{2n,0}_{[-1+n^\ast,-1+n^\ast][0,0,0]} \;,
\nln \Yboxdim4pt {\cal Z}_{\yng(1,1)}\eq
\sum_{n=0}^\infty\mult^{2n+1,0}_{[-\frac{1}{2}+n^\ast,-\frac{1}{2}+n^\ast][0,0,0]}
\;, \nln \Yboxdim4pt {\cal Z}_{\yng(3)} \eq \sum_{n=0}^\infty
c_n\left[
\mult^{1+n,0}_{[-1+\frac{1}{2}n^\ast,-1+\frac{1}{2}n^\ast][0,1,0]}
+ \bigbrk{\mult^{\frac{11}{2}+n,
\frac{1}{2}}_{[\frac{3}{2}+\frac{1}{2}n^\ast,1+\frac{1}{2}n^\ast][0,0,1]}
+\mbox{h.c.}}\right] \nl +\sum_{m=0}^\infty\sum_{n=0}^\infty
c_n\left[
\mult^{4+4m+n,1}_{[1+2m+\frac{1}{2}n^\ast,\frac{1}{2}n][0,0,0]}
+\mult^{9+4m+n,1}_{[\frac{7}{2}+2m+\frac{1}{2}n^\ast,
\frac{3}{2}+\frac{1}{2}n][0,0,0]} +\mbox{h.c.}\right] \;, \nln
\Yboxdim4pt {\cal Z}_{\yng(2,1)} \eq \sum_{n=0}^\infty d_n\left[
\mult^{2+n,0}_{[\frac12 n-\frac12^\ast,
\frac{1}{2}n-\frac12^\ast][0,1,0]} +\bigbrk{ \mult^{
\frac{7}{2}+n, \frac{1}{2}}_{[\frac12+\frac{1}{2}n^\ast,
\frac{1}{2}n^\ast][0,0,1]} +\mbox{h.c}}\right] \nl
+\sum_{m=0}^\infty\sum_{n=0}^\infty d_n\left[
\mult^{5+4m+n,1}_{[\frac32+2m+\frac{1}{2}n^\ast,\frac12+\frac{1}{2}n][0,0,0]}
+ \mult^{7+4m+n,1}_{[\frac52+2m+\frac{1}{2}n^\ast,
\frac{1}{2}+\frac{1}{2}n][0,0,0]} +\mbox{h.c.}\right]\;, \nln
\Yboxdim4pt {\cal Z}_{\yng(1,1,1)} \eq \sum_{n=0}^\infty c_n\left[
\mult^{4+n,0}_{[\frac{1}{2}+\frac{1}{2}n^\ast,
\frac{1}{2}+\frac{1}{2}n^\ast][0,1,0]} +\bigbrk{ \mult^{
\frac{5}{2}+n, \frac{1}{2}}_{[\frac{1}{2}n^\ast,-
\frac{1}{2}+\frac{1}{2}n^\ast][0,0,1]} +\mbox{h.c}}\right] \nl
+\sum_{m=0}^\infty\sum_{n=0}^\infty c_n\left[
\mult^{6+4m+n,1}_{[2+2m+\frac{1}{2}n^\ast,\frac{1}{2}n][0,0,0]} +
\mult^{7+4m+n,1}_{[\frac{5}{2}+2m+\frac{1}{2}n^\ast,
\frac{3}{2}+\frac{1}{2}n][0,0,0]} +\mbox{h.c.}\right]\;.
\label{456} \>

In table \ref{hscont} we rewrite the content of $\alhs(2,2|4)$
currents  in the symmetric doubleton.
\begin{table}[h]\centering
$\begin{array}{|c|c|} \hline
\alSU(4) &  \alSU(2)\times \alSU(2)   \\[2mm] \hline
{\bf 1} &    \sum_{r=-2}^2 [\ell-\ft{r}{2},\ell-\ft{r}{2}]+[\ell-1,\ell-1]+[\ell-1,\ell+1]  \\[2mm]
{\bf 4} &
[\ell-\ft{1}{2},\ell-1]+[\ell-\ft{1}{2},\ell]+[\ell-1,\ell+\ft{1}{2}]
+[\ell+1,\ell+\ft{1}{2}]\\[2mm]
{\bf 6} &  [\ell-1,\ell]+  [\ell+\ft{1}{2},\ell-\ft{1}{2}]+[\ell,\ell+1]  \\[2mm]
{\bf 10} &   [\ell+\ft{1}{2},\ell-\ft{1}{2}] \\[2mm]
{\bf 15} &   [\ell-\ft{1}{2},\ell-\ft{1}{2}]+[\ell,\ell+1] +[\ell,\ell] \\[2mm]
{\bf 20} &   [\ell-\ft{1}{2},\ell]  \\[2mm]
 {\bf 20}^\prime &  [\ell,\ell]\\[2mm]
 \hline
\end{array}$
\caption{Content of
${\mult}_{[000][\ell-1^*,\ell-1^*]}^{2\ell,0}$ for $\ell\geq 2$.}
\label{hscont}
\end{table}

\newpage

%\bibliographystyle{nb}
%\bibliography{hstot}

\end{document}